\documentclass[12pt,preprint]{aastex}
\usepackage{graphicx}
\usepackage{rotating}
\usepackage{hyperref}
\usepackage{amsmath}
\usepackage{lscape}
\usepackage{color}
\usepackage[english]{babel}

\newcommand{\ai}{{\it ab initio}}

\newcommand{\cm}{cm$^{-1}$}

\def\a0{{$a_{\rm 0}$}}

\def\Xstate{{X~$^{1}\Sigma_{\rm g}^{+}$}}
\def\Astate{{A~$^{1}\Pi_{\rm u}$}}
\def\Dstate{{D~$^{1}\Sigma_{\rm u}^{+}$}}
\def\Cstate{{C~$^{1}\Pi_{\rm g}$}}
\def\Bstate{{B~$^{1}\Delta_{\rm g}$}}
\def\Estate{{E~$^{1}\Sigma_{\rm g}^{+}$}}
\def\astate{{a~$^{3}\Pi_{\rm u}$}}
\def\bstate{{b~$^{3}\Sigma_{\rm g}^{-}$}}
\def\cstate{{c~$^{3}\Sigma_{\rm u}^{+}$}}
\def\dstate{{d~$^{3}\Pi_{\rm g}$}}
\def\estate{{e~$^{3}\Pi_{\rm g}$}}

\newcommand{\Duo}{{\sc Duo}}

\renewcommand{\baselinestretch}{1.1}

\shorttitle{Energy levels of C$_2$}
\shortauthors{Furtenbacher et al.}
\slugcomment{To appear in Astrophys. J. Suppl.}

\begin{document}

\title{Experimental energy levels and
partition function of the $^{12}$C$_2$ molecule}

\author{Tibor Furtenbacher, Istv\'an Szab\'o  and Attila G. Cs\'asz\'ar}
\affil{Institute  of Chemistry, Lor\'and E\"otv\"os University and
MTA-ELTE Complex Chemical Systems Research Group,
H-1518 Budapest 112, P.O. Box 32, Hungary}
\email{csaszar@chem.elte.hu}

\author{Peter F. Bernath}
\affil{Department of Chemistry, Old Dominion University, Norfolk, VA, U.S.A.}

\author{Sergei N. Yurchenko and Jonathan Tennyson}
\affil{Department of Physics and Astronomy, University College London,
 London WC1E 6BT, United Kingdom}
\email{j.tennyson@ucl.ac.uk}

\date{\today}
\begin{abstract}
The carbon dimer, the $^{12}$C$_2$ molecule, is ubiquitous in astronomical environments.
Experimental-quality rovibronic energy levels are reported for $^{12}$C$_2$,
based on rovibronic transitions measured for and among its singlet,
triplet, and quintet electronic states, reported in 42 publications.
The determination
utilizes the Measured Active Rotational-Vibrational Energy Levels (MARVEL) technique.
The 23,343 transitions measured experimentally and validated within this study
determine 5,699 rovibronic energy levels,
1,325, 4,309, and 65 levels for the singlet, triplet, and quintet states investigated,
respectively.
The MARVEL analysis provides rovibronic energies for six singlet,
six triplet, and two quintet electronic states.
For example, the lowest measurable energy level of the \astate\   state,
corresponding to the $J=2$ total angular momentum quantum number and the
$F_1$ spin-multiplet component, is 603.817(5) \cm.
This well-determined energy difference should facilitate observations of singlet--triplet
intercombination lines which are thought to occur in the interstellar
medium and comets.
The large number of highly accurate and clearly labeled transitions
that can be derived by combining MARVEL energy levels with computed
temperature-dependent intensities should help a number of astrophysical
observations as well as corresponding laboratory measurements.
The experimental rovibronic energy levels, augmented, where needed,
with {\it ab initio} variational ones based on empirically adjusted and spin-orbit coupled
potential energy curves obtained using the \Duo\ code,
are used to obtain a highly accurate partition function,
and related thermodynamic data, for $^{12}$C$_2$ up to 4,000 K.
\end{abstract}
\keywords{C$_2$, MARVEL, rovibronic transitions,
rovibronic energy levels, internal partition function}

\renewcommand{\baselinestretch}{1.0}
\begin{deluxetable}{llll}%[!htbp]
\tablecaption{Singlet, triplet, and quintet band systems of $^{12}$C$_2$ for which
rovibronic transitions have been reported in the literature.
Band systems printed in italics, due to reasons given below, are not considered in
this paper.}
%\resizebox{\columnwidth}{!}{%
\tablewidth{0pt}
\tablehead{
\colhead{multiplicity}&\colhead{band system} & \colhead{transition} & \colhead{original detection}}
\startdata
singlet & Phillips                          & \Astate $-$ \Xstate   & \citet{48Phillips.C2a}\\
        & Mulliken                          & \Dstate $-$ \Xstate   & \citet{39Landsver.C2} \\
        & {\it Herzberg F}                  & F~$^{1}\Pi_{\rm u}-$\Xstate&\citet{69HeLaMa}  \\
        & Bernath B                         & \Bstate $-$ \Astate   & \citet{88DoNiBeb.C2}    \\
        & Bernath B${'}$                   & B${'}$~$^{1}\Sigma_{\rm g}^{+}-$\Astate\  & \citet{88DoNiBeb.C2}  \\
        & {\it Deslandres--d'Azambuja}      & \Cstate $-$ \Astate   & \citet{1905DeDa.C2}\\
       	& {\it Messerle--Krauss}            & C${'}$~$^{1}\Pi_{\rm g} -$ \Astate\ & \citet{67MeKrxx.C2}    \\
        & Freymark                          & \Estate$-$ \Astate    & \citet{51Freymark.C2}\\
        & {\it Goodwin--Cool A}             & 1~$^1\Delta_{\rm u} -$ A $^1\Pi_{\rm u}$ & \citet{88GoCoa}    \\
	      & {\it Goodwin--Cool B}             & 1~$^1\Delta_{\rm u} -$ B $^1\Delta_{\rm g}$& \citet{88GoCoa}    \\
triplet & Ballik--Ramsay                    & \bstate $-$ \astate   & \citet{58BaRa}  \\
        & Swan                              & \dstate $-$ \astate   & \citet{1857Swan.C2} \\
        & Fox--Herzberg                     & e~$^{3}\Pi_{\rm g}-$ \astate & \citet{37FoHe.C2} \\
        & {\it Herzberg f}& f~$^{3}\Sigma_{\rm g}^{-}-$ \astate & \citet{69HeLaMa} \\
        & {\it Herzberg g}& g~$^{3}\Delta_{\rm g}-$ \astate  & \citet{69HeLaMa}\\
        & Krechkivska--Schmidt              & 4 $^3\Pi_{\rm g} - $ a $^3\Pi_{\rm u}$& \citet{15KrBaTrNa}\\
        & Duck                              & \dstate $-$ \cstate   &\citet{06KoReMoNa.C2} \\
        & {\it Kable--Schmidt}              & e~$^{3}\Pi_{\rm g} -$ \cstate & \citet{09NaJoPaRe.C2}\\
quintet & Radi--Bornhauser                  & $1~^{5}\Pi_{\rm u}-1~^{5}\Pi_{\rm g}$& \citet{11BoSyKnGe.C2} \\
intercombination& triplet-singlet           & \astate  $-$ \Xstate        & \citet{15ChKaBeTa.C2} \\
        & quintet-triplet           &$1~^{5}\Pi_{\rm g} -$ \astate& \citet{11BoSyKnGe.C2} \\
        & singlet-triplet             & \Astate $-$ \bstate         & \citet{15ChKaBeTa.C2}\\
\\
\enddata
\label{tab:BANDS}
\end{deluxetable}

\begin{figure}
\includegraphics[height=160mm]{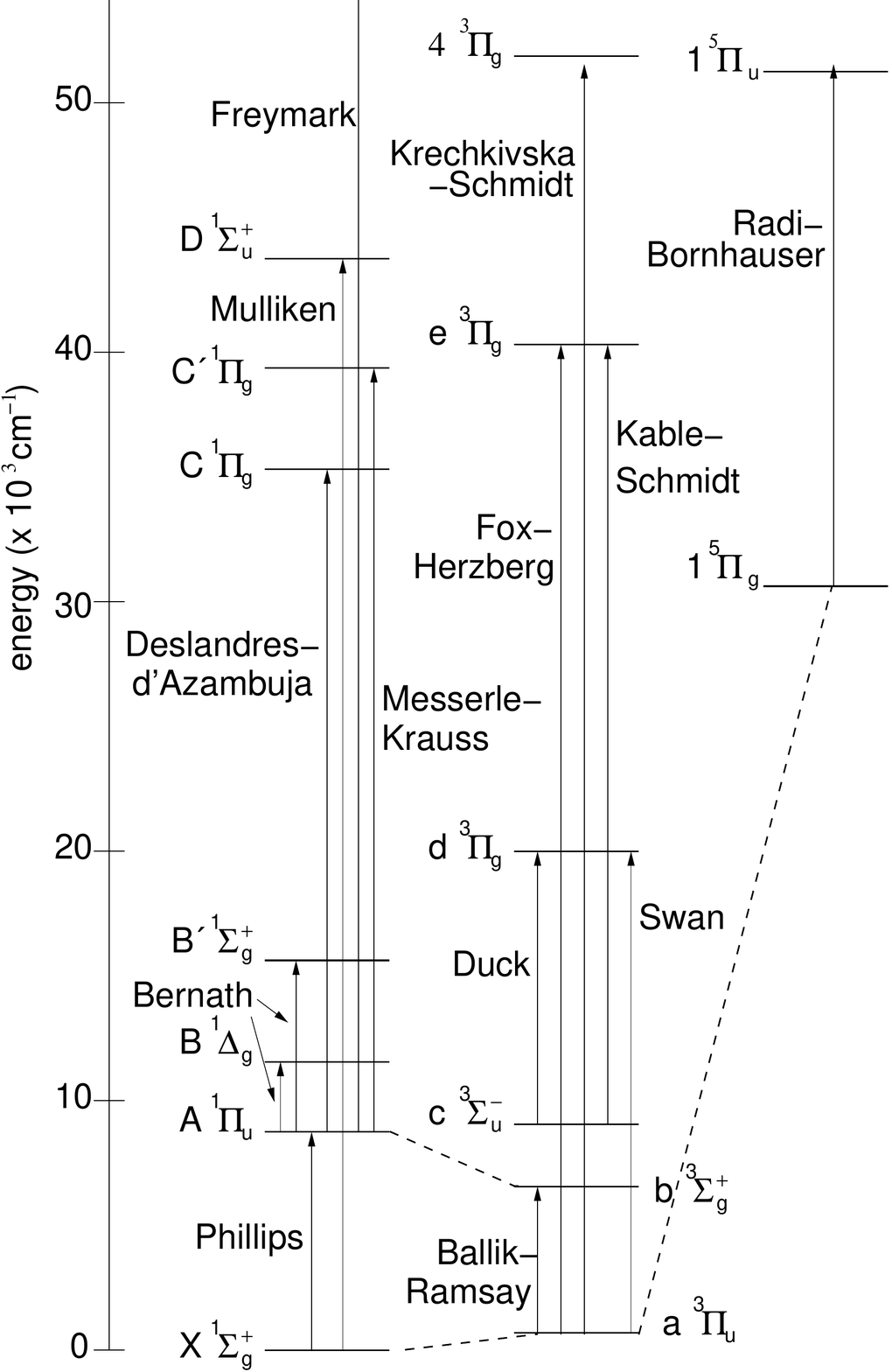}
\caption{The band system of $^{12}$C$_2$ showing the bands considered in this work.
The dashed lines represent observed but unnamed intercombination bands.}
\label{fig:C2bands}
\end{figure}

\renewcommand{\baselinestretch}{1.25}
\section{Introduction}
The rovibronic spectra of $^{12}$C$_2$, involving singlet, triplet, and quintet
electronic states, are rich and complex in features as well as in anomalies.
This is true despite the fact that $^{12}$C$_2$, hereafter simply called C$_2$,
is a homonuclear diatomic molecule containing only 8 valence electrons.
There are 62 electronic states corresponding to the six possible separated-atom limits
formed by different pairs of the C($^3$P), C($^1$D), and C($^1$S) atoms.
18 of these states, six singlet, six triplet, and six quintet ones,
correspond to the C($^3$P) + C($^3$P) asymptote.
As summarized in Table~1 and Figure~1,
the measured rovibronic spectroscopy of C$_2$ is characterized by
19 known band systems, and a further three intercombination transitions,
which cover the spectral range $0-55,000$ \cm\   and now all three spin states.
Due to the lack of detailed transition data, see section 2.4 for details,
only 14 band systems are considered in the present study.
All the states considered, except the \estate\  and the 4~$^{3}\Pi_{\rm g}$ states,
share the same C($^3$P) + C($^3$P) separated-atom limit.
The present study was initiated as, rather unusually,
at least six of these bands have been used for astronomical
observations, with a further one having been proposed for such use.

There are numerous astronomical observations of C$_2$ spectra with a
multitude of important applications for astrophysics.  C$_2$ spectra
have been observed in comets \citep{68MaDe}, in high-temperature stars
\citep{70Vartya}, translucent clouds \citep{07SoWeThYo}, and in the
low-temperature interstellar medium \citep{77SoLu}.  High-quality
studies of rotational and spatial distributions of C$_2$ in comets are
available \citep{78Lambert,90LaShDaAr}.  The astronomical observations
and their modelling are supported by a number of laboratory studies
\citep{85WoLe,88UrBaJa.C2,91JaBaUr,91BaUrJa,95BlLiSo.C2,96BlLiSo.C2}.
To make maximum use of the observations one needs accurate rovibronic
energy level information, the principal topic of the present paper.

Both the Swan \citep{43Swings,89GrVaBl,90LaShDaAr,00RoHiBu} and the
Deslandres--d'Azambuja \citep{89GrVaBl} bands have been observed in
fluorescence from comets.  Indeed, two singlet, \Xstate\ and \Astate,
and four triplet, \astate, \bstate, \cstate, and d~$^{3}\Pi_{\rm g}$,
electronic states are needed to model C$_2$ emission observed in
comets.  Two spin-forbidden transition systems, \astate $\rightarrow$
\Xstate\ and \cstate $\rightarrow$ \Xstate, are needed to explain the
observed intensities in the Swan band \citep{00RoHiBu}.

The astronomical observations include solar spectra, where C$_2$ forms an
important component of the photospheric carbon abundance \citep{05AsGrSaPr}.
So, for example, C$_2$ can be observed in the Sun's photosphere at
visible wavelengths using the Swan band \citep{05AsGrSaPr} and in the
infrared via the Phillips and Ballik--Ramsay  bands \citep{82BrDeGr}.
Transitions in the Swan band have also been observed in peculiar white
dwarfs \citep{08HaMa.C2,10Kowalski} and the coronae borealis star V coronae
australis \citep{08RaLa}, while the Phillips band is prominent,
for example, in the carbon star HD19557 \citep{83GoBrCo.C2}.
Transitions in the Ballik--Ramsay band have also been observed in carbon
stars \citep{90Goorvitc}.

Interstellar C$_2$ has been observed via the infrared Phillips band,
for example in the Perseus molecular complex
\citep{79Hobbs.C2,95LaShFe.C2,11Iglesias} and toward the Cyg {OB2}
association \citep{01GrBlYa.C2}.
Emission features from the Swan band
have been observed in the Red Rectangle \citep{10WeRoLi}. \citet{07SoWeThYo}
observed absorbotion in the Phillips, Mulliken and Herzberg F bands in
their study of translucent clouds by the simultaneous use of observations
from both space and ground-based telescopes.

The rovibrational manifolds of the \Xstate\   and \astate\  states
strongly overlap as the corresponding electronic excitation
energy is less than half of the vibrational spacing in either state.
\citet{86LeRo} suggested that long-wavelength transitions between
levels in the \astate\   -- \Xstate\  band should provide
a good method for monitoring interstellar C$_2$. \citet{98RoLaMoCl}
attempted to observe such lines in the Hale--Bopp comet, without success.
The accurate rovibronic energy levels presented in this paper
can be used to provide accurate transitions for such features.

As for the other bands,
at ultraviolet wavelengths the International Ultraviolet Explorer
(IUE) was used to observe C$_2$ towards X Persei via the
Herzberg F band  \citep{84Lien}.
The Hubble Space Telescope also has been used to monitor C$_2$ absorption in
the ultraviolet using both the Mulliken and the Herzberg F
bands \citep{95LaShFe.C2,12HuShFe}.

Finally, we note that all of the bands mentioned are important in
laboratory plasmas \citep{90ChMa,98DuStSu,11Nemes}
and flames \citep{65BeNixx,79AmChMa.C2,98BrHaKoCr,99LlEw,08GoCh,05SmPaSc}.
For example, there are particular vibrational bands of the Swan system known as the
High Pressure (HP) bands \citep{46Herzberg,87LiBr,94CaDo} which can be
prominent in such environments at atmospheric pressure.

Studies of the spectra, and thus the band systems, of C$_2$ date all the way back
to \citet{1802Wo.C2} and the dawn of spectroscopy,
when C$_2$ emissions were first observed in flames.
Detailed studies by \citet{1857Swan.C2} of the most prominent band of C$_2$
also predate the development of the quantum mechanical tools required to interpret
the spectroscopic results obtained.
Over the following century and a half,
a number of other bands have been detected and studied.
Indeed, four  band systems have been identified during the last
decade \citep{06KoReMoNa.C2,09NaJoPaRe.C2,11BoSyKnGe.C2,15KrBaTrNa}.
Table~\ref{tab:BANDS} lists the known band systems of C$_2$ relevant for the present study.

A large number of spectroscopic measurements exist for the different
band systems of C$_2$.  All the studies which contain primary measured
transitions at a reasonable level of accuracy are considered during
the present analysis.  The experimental papers found and analysed by
us are listed here based on the band systems: Phillips
\citep{48Phillips.C2a,63BaRab.C2,77ChMaMa.C2,82ErLaMa.C2,88DaPhRaAb,88DaAbPh.C2,88DoNiBea.C2,04ChYeWoLi,06PeSi,13NaEn.C2,15ChKaBeTa.C2},
Mulliken \citep{39Landsver.C2,39Mulliken.C2}, Bernath~B
\citep{88DoNiBeb.C2,91BaUrJa}, Bernath~B$'$ \citep{88DoNiBeb.C2},
Freymark \citep{51Freymark.C2,96BlLiSo.C2}, Ballik--Ramsay
\citep{63BaRaa.C2,75Veseth,88DaAbSa.C2,79AmChMa.C2,85RoWaMiVe.C2,85YaCuMeCa,06PeSi,11BoSyKnGe.C2,15ChKaBeTa.C2},
Swan
\citep{1857Swan.C2,1916Raffety,63CaGi,68MeMe,68Phillips,68PhDaxx.C2,79BrHa,83Amiot,85CuSa,85SuSaHi.C2,88UrBaJa.C2,90ChMa,97KaHuEw,99LlEw,05SmPaSc,06KoReMoNa.C2},
Fox--Herzberg \citep{37FoHe.C2}, Deslandres--d'Azambuja
\citep{1905DeDa.C2,85AnBoPe.C2,87VaHe,88UrBaJa.C2}, Messerle--Krauss
\citep{67MeKrxx.C2}, \\Goodwin--Cool \citep{88GoCoa,88GoCob,89GoCo},
Duck \citep{87VaHe,06KoReMoNa.C2} , Krechkivska-Schmidt
\citep{15KrBaTrNa,16KrBaWeNa} and Radi-Bornhauser \citep{15BoMaGo}, as
well as various spin-changing intercombination bands
\citep{11BoSyKnGe.C2,15ChKaBeTa.C2}.

It is interesting to note that for a long time a triplet state, now named \astate,
was believed to be the lowest electronic state of C$_2$ and
not a singlet state, the true ground electronic state, \Xstate.
The source of the confusion, detailed in \citet{59BaRa.C2}, is that the energy difference
between the two electronic states is only about one half of the vibrational fundamental
of either state so that the Swan system can be seen in absorption in many sources.
The spectroscopy and the spectroscopic constants of C$_2$ have been
the subject of several reviews \citep{77HuHe,89WeVa,92Martin,98VaSa}.
\citet{77HuHe} reported the analysis
of seven singlet and seven triplet states.
\citet{89WeVa} reviewed the then available
experimental and theoretical results.
The spectroscopic and kinetic properties of C$_2$,
including  23 electronic states studied prior to 1992,
were reviewed by \citet{92Martin}.
\citet{98VaSa} reviewed the spectroscopy of small carbon clusters which,
once again, covered the extensive spectroscopic literature available for C$_2$.
The present paper
surveys all rovibronically-resolved measurements made up to the end of 2015.

A large number of {\it ab initio} computations have been performed on C$_2$
\citep{79KiLi.C2,87BaLa.C2,92WaBart.C2,00BoVoHa.C2,01BrGrein.C2,01MuDaLiDu,04AbSh.C2,05ShPixx.C2,07KoBaSc.C2,11BoClThAl,11ScBa,11JiWi,12AnCiPa,13BoThRu.C2,14BoThRuWi,15KrBaTrNa,16KrBaWeNa}.
Notable among the sophisticated electronic structure computations
are those of Bacskay
\citep{07KoBaSc.C2,11ScBa,15KrBaTrNa,16KrBaWeNa};
for example, they led to the observation of the Duck band
and enabled the identification of the Krechkivska--Schmidt band system.
\citet{14BoThRuWi} pointed out that
``at the present state of the art, theoretical PECs that reproduce
the rotational-vibrational
levels to spectroscopic accuracy (1 \cm) or near spectroscopic accuracy (10 \cm)
are considered highly accurate.''
Reproduction of measured electronic excitation energies of C$_2$ appears to be
even more problematic, an accuracy of a few hundred \cm\   seems to be the norm.
Thus, it is still of interest to perform accurate quantum chemical computations
on C$_2$ and its rovibronic states.
Results of a preliminary, first-principles analysis are reported here
({\it vide infra}),
used in particular for checking the experimental transitions and energy levels.

The main body of  the present work is a
MARVEL (Measured Active Rotational-Vibrational Energy Levels)
\citep{jt412,12FuCs} analysis of the measured rovibronic states of C$_2$.
Our analysis has been in progress for four years but  a real breakthrough came
with the study of \citet{15ChKaBeTa.C2}, who observed 16 forbidden
transitions between singlet and triplet states.
This study coupled,
for the first time, the singlet and triplet components of the observed
spectroscopic network (SN) \citep{11CsFu,11FuCs,16CsFuAr} of C$_2$, allowing a
much improved analysis of its rovibronic energy level structure.
We analyze all the known bands of C$_2$ with the
exception of the five VUV bands due to \citet{69HeLaMa} and
\citet{88GoCoa,88GoCob,89GoCo}.
These bands  involve upper energy levels which arise from a
single experiment and are too high for
us to be able to independently validate them.

Finally, when a complete set of accurate energy levels are available for a
molecule, they can be used, via the direct summation technique,
to compute accurate ideal-gas thermodynamic functions, most importantly the
high-temperature internal partition function, $Q_{\rm int}(T)$.
We do this here for C$_2$, complementing and improving several previous
efforts \citep{60Altman,61Clementi,81Irwin.partfunc,84SaTaxx.partfunc,85RoMaVe.partfunc,90Gurvich}
and arrive at very precise and accurate values
for $Q_{\rm int}(T)$ of C$_2$ up to 4,000 K.

%\newpage
\begin{figure}
\includegraphics[height=150mm]{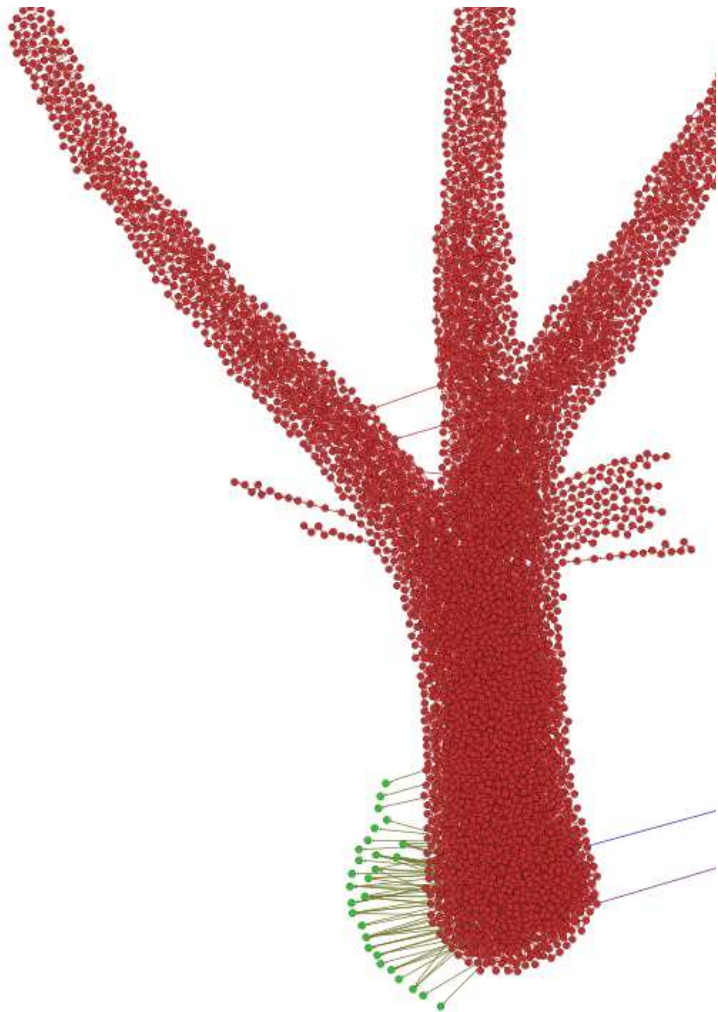}
\caption{
Representation of the experimental spectroscopic network built in this study for C$_2$.
The blue, red, and green dots correspond to the singlet, triplet,
and quintet rovibronic states, respectively.
The dots represent energy levels and only some of the transitions are visible,
especially those connecting the singlet and the triplet as well as the triplet
and the quintet rovibrational states.}
\label{f:CNM}
\end{figure}

\section{Methodological details}

\subsection{Experimental spectroscopic network of C$_2$}
The MARVEL \citep{jt412,12FuCs} procedure and code is used in this study to
obtain rovibronic energies of C$_2$ by inverting all assigned experimental
rovibronic transitions available in the literature for this molecule.
%analyze all experimental measured rovibronic transitions of C$_2$,
MARVEL is based on the concept of spectroscopic networks (SN) \citep{11CsFu,11FuCs,16CsFuAr}:
SNs are large, finite, weighted, rooted graphs, where the vertices are
discrete energy levels (with associated uncertainties),
the edges are transitions (with measured uncertainties),
and in a simple picture the weights are provided by the transition intensities.
No weights are considered during the present study.
Within the experimental SN of a molecule there can be several rooted
components and several floating ones.
For many molecules the rooted components belong to ortho and para nuclear
spin isomers.
However, the nuclear spin of $^{12}$C is zero; thus, $^{12}$C$_2$ has only one
nuclear spin isomer and one root for its lowest electronic state, \Xstate.

Transitions between electronic states of different spin multiplicity are spin forbidden;
thus, until 2011 no lines were measured and assigned experimentally between the
singlet and triplet and the triplet and quintet manifold of states of C$_2$.
First, \citet{11BoSyKnGe.C2} observed
transitions linking triplet and quintet electronic states,
leading to the first observation and characterization of a quintet
band of C$_2$.
% \citep{11BoSyKnGe.C2}.
Later, \citet{15ChKaBeTa.C2} managed to identify 16 spin-forbidden transitions
between singlet and triplet states.
These spin-forbidden transitions proved to be particularly important during
the MARVEL analysis of the experimental spectra of C$_2$.
Using these forbidden transitions the experimental SN of C$_2$
simplifies as now it contains only one principal component (PC).
The label of the root of the PC is \{0 + \Xstate\ 0 $F_1$ $e$\},
%\textcolor[rgb]{1.00,0.00,0.00}{(Has e/f parity been inserted correctly? This should have been e and I changed it from f. PB)},
for the notation employed for the label see subsection \ref{s:QN}.

The measured SN of C$_2$ is shown pictorially in Figure 2, the singlet,
triplet, and quintet energy levels are indicated with different colors.
Since Figure 2 is a particular representation of a network ($i.e.$, a graph),
the arrangement of the nodes (energy levels) and links (transitions) is
arbitrary, but clearly displays several important characteristics of the SN of C$_2$.
Figure 2 vividly shows, for example, how weakly the singlet rovibrational energy levels
are connected to the triplet core.
%It must also be noted that due to the nature of the experiments performed the
%rovibronic transitions involve several electronic states even in the
%branches of the SN, as also visible in the figure.

\subsection{MARVEL}
During a MARVEL analysis we simultaneously process all the available
assigned and labelled experimental transitions.
The energy levels are obtained from the set of transitions via a weighted
linear least-squares inversion protocol.
As the MARVEL technique has been employed to determine experimental-quality
energy levels of nine isotopologues of water \citep{jt454,jt482,jt539,jt576,jt562},
of three isotopologues of H$_3^+$ \citep{13FuSzFaCs,13FuSzMaFa},
as well as of ammonia \citep{jt608} and ketene \citep{11FaMaFuMi},
the interested reader is referred to these publications for details
about the different stages of a MARVEL analysis.

Since C$_2$ has both regular and inverted triplet states
(see the next subsection for details), for example the \astate\   and
\dstate\   of C$_2$ are inverted, during the MARVEL analysis it was checked
whether the labels of all experimental transitions follow the same convention.

At the beginning, the MARVEL analysis of the spectra of C$_2$ was complicated by the
fact that there have been no truly high-accuracy variational or effective Hamiltonian
rovibronic energy levels available for this molecule at higher energies.
This situation greatly improved by the first-principles \Duo\  \citep{jt609}
analysis performed as part of this study ({\it vide infra}), allowing at least a
preliminary validation of the MARVEL levels up to 35,000 \cm.
%At the end, we divided the MARVEL rovibronic energy levels of C$_2$
%into two groups: fully supported and conditionally supported energy levels.
%It is important to emphasize that ``conditional support''
%does not mean that the energy level is wrong, it only means that
%we could not check its value by DUO and validate it via cycles
%during the MARVEL analysis.

\subsection{Labels and quantum numbers}
\label{s:QN}

For MARVEL to work properly one needs appropriate and unique labels.
The label used in the present study for a rovibronic energy level
is built up from information concerning the uncoupled electronic state,
the vibrational and rotational quantum numbers, and the symmetry of the
rovibronic state.

%The Hund's case quantum numbers are used. LAMBDA IS EXCLUDED SINCE STATE COVERS IT,
%BUT REDUNDANT e/f, g/u ADDED FOR CONVENIENCE;

The label finally chosen for each energy level,
% contains the following information:
%$  \{ J, \,  +/- ,  \,  {\rm `State'}, \,  v, \,  F$_i$, \}$,
$  \{ J, +/-, {\rm state}, v, F_i, e/f \}$,
%This label
fully characterizes the rovibronic states of a homonuclear diatomic
molecule, such as C$_2$, but contains redundant
information \citep{BrownCarrington,BrionField,16Bernat.method}.
Here $J \geq 0$ is the quantum number corresponding to the total angular
momentum (an integer value) characterizing the state,
$+/-$ labels the total state parity, $+1/-1$, in terms of the laboratory-fixed
inversion operator $E^*$ \citep{16Bernat.method},
`state' is the customary term symbol of the electronic state before
spin-orbit coupling is taken into account
({\it e.g.}, X~$^{1}\Sigma^{+}_{\rm g}$ and b~$^{3}\Sigma^{-}_{\rm g}$),
and $v$ is the vibrational quantum number.
%and $F_i$ denotes the spin multiplet components.
The $F_i$, $i=1,2,...,2S+1$, label denotes the spin multiplet components:
for singlet states $i=1$, {\it i.e.}, $F_i \equiv F_1$,
%\red{SERGEY: WE USE 0, BUT I SUGGEST TO CHANGE TO 1},
for triplet states $i=1,2,3$,
which corresponds to the standard spectroscopic notation
$F_1, F_2$, and $F_3$ ($F_1$, $F_2$, $F_3$ refer to levels with $J=N+1,N,N-1$,
where $N$ is the quantum number corresponding to the angular momentum exclusive
of nuclear and electron spin and it is usually not a good quantum number),
and for quintet states $i=1,2,3,4,5$  \citep{73Whiting.NASA}.
The convention \citep{16Bernat.method} is such that whether the state is
regular ($A>0$) or inverted ($A<0$), the energy order is always
$F_3>F_2>F_1$ for the triplet states.
The $g$ and $u$  subscript in the `state' label indicate if the symmetry
of the electronic state is `gerade' (positive) or `ungerade' (negative)
in terms of the molecule-fixed inversion operator, $i$ \citep{16Bernat.method}.
The $+/-$ superscripts within the $\Sigma^{+/-}$ states indicate the parity
of the electronic component with respect to vertical reflection, $\sigma_v$,
in the molecular frame \citep{16Bernat.method}.
The rotationless parity $e/f$ is a widely used alternative to the total parity $+/-$,
this redundant information is included in our MARVEL label to help experimentalists.
For C$_2$ the allowed combination of the parity $+/-$ and the label g/u is
($+$,g) and ($-$,u).
The rotationless parity $e/f$ is then obtained as follows:
%connected to the total state parity via the following standard rule:
%the $e$ and $f$ levels have the total parity $(-1)^{J}$ and $(-1)^{J+1}$, respectively.
(a) for the $e$ states,
the parity ($+1$ or $-1$) can be recovered as $(-1)^{J}$, {\it i.e.},
for the even values of $J$ the parity of the $e$ states is $+1$,
while for odd $J$s it is $-1$;
and (b) for the $f$ states,
the parity is recovered as $(-1)^{J+1}$, {\it i.e.},
for even $J$s the parity is $-1$, and for odd $J$s it is $+1$.

The rigorous electric dipole selection rules for the rovibronic transitions are
\begin{eqnarray}
 J^{\prime\prime} - J^{\prime}  = \pm 1, 0, \;\; &,& \;\; 0 \not\leftrightarrow 0, \\
  + \leftrightarrow -  \;\; &,& \;\;  g \leftrightarrow u.
\end{eqnarray}

%Finally, we note that the $+/-$ and the ${\rm g}/{\rm u}$ symmetry labels of the $D_{\infty h}$ point group correlate with the
%$A_1$, $A_2$, $B_1$, and $B_2$ irreducible representations of $C_{2v}$ as follows:
%\begin{eqnarray}
%A_1 &\leftrightarrow& +,g, \\
%A_2 &\leftrightarrow& -,u,  \\
%B_1 &\leftrightarrow& +,u, \\
%B_2 &\leftrightarrow& -,g.
%A_1 &\leftrightarrow& \Sigma^+_g, \\
%A_2 &\leftrightarrow& \Sigma^-_u,  \\
%B_1 &\leftrightarrow& \Sigma^-_g, \\
%B_2 &\leftrightarrow& \Sigma^+_u,
%\end{eqnarray}
%where $\Sigma^{+/-}_{g/u}$ is a \red{reducible} \blue{PFB: irreducible? How does this work?
%representation of $D_{\infty h}$.
%The corresponding selection rules are
%%$$
%\begin{eqnarray}
%A_1 \leftrightarrow A_2 \;\; {\rm and} \;\; B_1 \leftrightarrow B_2.
%\end{eqnarray}
%$$
As mentioned, the nuclear spin statistical weights $g_{\rm ns}$ of the nuclear
spin-zero $^{12}$C$_2$ molecule
for the $(+,g)$ and $(-,u)$ states is 1,
while the  $(-,g)$ and $(+,u)$ states have $g_{\rm ns} = 0$.
Therefore, the latter states do not appear in spectroscopic experiments on $^{12}$C$_2$.
The MARVEL input set of measured transitions was checked for a corresponding labeling
error.

\renewcommand{\baselinestretch}{1.0}
\begin{deluxetable}{l l c l c l}%[!htbp]
\tablecaption{Experimental transitions available in the literature for
several band systems of $^{12}$C$_2$ and their overall characteristics,
including the number of measured (A) and validated (V) transitions (Trans.).
Comments are given in section~\ref{com2}.
%the Phillips ($A^{1}\Pi_{u}-X^{1}\Sigma_{g}^{+}$),
%Ballik--Ramsay ($b^{3}\Sigma_{g}^{-}-a^{3}\Pi_{u}$), Swan ($d^{3}\Pi_{g}-a^{3}\Pi_{u}$),
%Bernath $B$ ($B^{1}\Delta_{g}-A^{1}\Pi_{u}$), Bernath $B'$ ($B'^{1}\Sigma_{g}^{+}-A^{1}\Pi_{u}$), and
%$d^{3}\Pi_{g}-c^{3}\Sigma_{u}^{+}$ systems.
}
\label{tab:EXP}
\tablewidth{0pt}
\tablehead{
\colhead{Tag}&    \colhead{Ref.}    & \colhead{Range(cm$^{-1}$)} &  \colhead{Method} &\colhead{Trans (A/V)}  & \colhead{Comments} }
\startdata

\multicolumn{2}{l}{Phillips \Astate$-$\Xstate}\\
15ChKaBeTa&\citet{15ChKaBeTa.C2}& $2372-8822$  &  FTS  & 319/318 & \\
88DaAbPh  & \citet{88DaAbPh.C2} & $4012-6031$  & D-FTS & 191/191 & \\
88DoNiBea & \citet{88DoNiBea.C2}& $4067-7565$  &  FTS  & 241/238 & \\
77ChMaMa  & \citet{77ChMaMa.C2} & $4371-11424$ & FTS   & 774/770 & \\
63BaRab   & \citet{63BaRab.C2}  & $6312-14469$ &       & 574/532 & \\
06PeSi    & \citet{06PeSi}      & $9405-9434$  &       &   8/8   & (2a)\\
04ChYeWoLi& \citet{04ChYeWoLi}  & $10719-14128$&       & 293/293 & (2b) \\
13NaEn    & \citet{13NaEn.C2}   &$13182-16825$ &  LIF  & 77/73   & \\
\\
\multicolumn{2}{l}{Mulliken \Dstate$-$\Xstate} \\
39Landsver &\citet{39Landsver.C2}&$43051-43473$ & CAE   & 171/167 & (2c) \\
97SoBlLiXu & \citet{97SoBlLiXu}  &$43062-43490$ & LIF   & 179/165  & (2c) \\
95BlLiSo   & \citet{95BlLiSo.C2} &$43224-43289$ & DL    &   9/8   & \\
\\
\multicolumn{2}{l}{Bernath \Bstate $-$ \Astate}\\
88DoNiBeb & \citet{88DoNiBeb.C2} & $1951-7152$  & FTS   & 507/507 & \\
16ChKaBeTa& \citet{16ChKaBeTa.C2}& $1900-8422$  & FTS   &1001/1001& \\
\\
\multicolumn{2}{l}{Bernath B$'$~$^{1}\Sigma_{\rm g}^+ -$\Astate}\\
88DoNiBeb  & \citet{88DoNiBeb.C2} & $5025-8365$  & FTS   &  237/237& \\
\\
\multicolumn{2}{l}{Freymark \Estate$-$\Astate} \\
97SoBlLiXu & \citet{97SoBlLiXu}  &$43510-43660$ & LIF   & 66/65   & (2c)\\
51Freymark &\citet{51Freymark.C2}&$45069-48772$ &       & 376/354 & (2d)\\
\\
\multicolumn{2}{l}{Ballik--Ramsay \bstate$-$\astate}\\
15ChKaBeTa &\citet{15ChKaBeTa.C2}& $2102-9404$  &  FTS  &3510/3507& \\
85YaCuMeCa & \citet{85YaCuMeCa}  & $3673-4040$  &       &356/352  & (2e)\\
85RoWaMiVe &\citet{85RoWaMiVe.C2}& $4643-8488$  &       &1309/1298& (2f)\\
79AmChMa   &\citet{79AmChMa.C2}  & $4856-9895$  &       &2168/2139& \\
88DaAbSa   & \citet{88DaAbSa.C2} & $4897-5706$  & D-FTS & 382/365 & \\
06PeSi     & \citet{06PeSi}      & $9388-9450$  &       & 80/80 & \\
11BoSyKnGe &\citet{11BoSyKnGe.C2} &$22991-23031$&       & 8/8   & \\
\\
\multicolumn{2}{l}{Swan    \dstate$-$\astate}\\
13NaEn     &  \citet{13NaEn.C2}  &$12596-23497$ &  LIF  &168/168  & \\
14NaEn     &  \citet{14NaEn.C2}  &$13673-13877$ &  LIF  & 150/141 & \\
13BoSyKnGe & \citet{13BoSyKnGe}  &$13847-13927$ &       & 23/23   & (2g)\\
13YeChWa   &  \citet{13YeChWa.C2}&$13849-14128$ &  HCD  & 276/273 & (2h) \\
07TaHiAm   &\citet{07TaHiAm.C2}  &$15149-23110$ &  FTS  &3853/3771& \\
48Phillips &\citet{48Phillips.C2a}&$16151-44722$&discharge&1181/1128&\\
02TaAm     & \citet{02TaAm.C2}   &$16877-17113$ &  HCD  & 356/352 & \\
03KaYaGuYu &\citet{03KaYaGuYu}   &$17731-17895$ &       & 153/150 & (2i) \\
85CuSa     & \citet{85CuSa}      &$17736-17941$ &Doppler-free&217/217& (2j)\\
94PrBe     & \citet{94PrBe.C2}   &$17915-21315$ &Jet cooled&39/39 & (2j) \\
83Amiot    &\citet{83Amiot}      &$19354-20191$ &   FTS & 347/346 & \\
99LlEw     &\citet{99LlEw}       &$19354-19511$ &       & 138/138 & \\
85SuSaHi   & \citet{85SuSaHi.C2} &$21101-21263$ &       & 194/194 & (2j) \\
10BoKnGe   & \citet{10BoKnGe.C2} &$21343-21427$ &       &  23/23  & (2j) \\
11BoSyKnGe &\citet{11BoSyKnGe.C2}&$21389-23019$ &       &  46/46  & \\
%(2j) \\
\\
\multicolumn{2}{l}{Fox--Herzberg  e~$^{3}\Pi_{\rm g}-$\astate}\\
86HaWi     & \citet{86HaWi}      &$33036-33492$ &       & 100/100  & (2k)  \\
49Phillips &\citet{49Phillips.C2}&$35092-42019$&discharge&1833/1664& \\
98BrHaKoCr & \citet{98BrHaKoCr}  &$40205-40339$ &       & 10/10   & (2l) \\
\\
\multicolumn{2}{l}{Duck \dstate$-$\cstate}\\
13ChYeWa  &  \citet{13ChYeWa.C2}&$12074-12499$ &  HCD  & 221/210 & \\
13NaEn    &  \citet{13NaEn.C2}  &$13030-16136$ &  LIF  & 513/513 & \\
14NaEn    &  \citet{14NaEn.C2}  &$13650-13889$ &  LIF  & 205/196 & \\
07JoNaRe  &  \citet{07JoNaRe.C2}&$15007-17080$ &  LIF  & 235/234 &  \\
\\
\multicolumn{2}{l}{Krechkivska--Schmidt $4^3\Pi_{\rm g}-$\astate} \\
15KrBaTrNa& \citet{15KrBaTrNa}  &$47921-48327$ & REMPI & 67/67   & \\
16KrBaWeNa& \citet{16KrBaWeNa}  & $46736$      &       &  1/1    & (2m)\\
\\
\multicolumn{2}{l}{Intercombination \astate$-$\Xstate}\\
15ChKaBeTa  &\citet{15ChKaBeTa.C2}&$3501-8306$   &  FTS  &  32/32  & \\
\\
\multicolumn{2}{l}{Intercombination \Astate$-$\bstate}\\
15ChKaBeTa  &\citet{15ChKaBeTa.C2}& $3940$       &  FTS  &  1/1    & \\
\\
\multicolumn{2}{l}{Intercombination  $1~^{5}\Pi_{\rm g} -$ \astate}\\
11BoSyKnGe &\citet{11BoSyKnGe.C2}&$21370-21447$ &  FWM  &  68/68  &  \\
\\
\multicolumn{2}{l}{Radi--Bornhauser $1^5\Pi_{\rm u}-1^5\Pi_{\rm g}$}  \\
15BoMaGo  & \citet{15BoMaGo}    &$21772-21839$ &  FWM & 57/57   & \\
\enddata

\tablecomments{CAE = carbon arc emission, D-FTS = discharge FTS, DL = dye laser, \\
FTS = Fourier Transform Spectroscopy, FWM = four-wave mixing, \\
HCD = Hollow-Cathode Discharge source, LIF = Laser Induced Fluorescence, \\
MD = Microwave Discharge, SJT = Supersonic Jet Technique, \\
%Trans = transitions, A = transitions available, V = transitions validated, \\
REMPI = resonance-enhanced multiphoton ionization.}
\end{deluxetable}

\renewcommand{\baselinestretch}{1.0}
\begin{deluxetable}{l l c l c l}%[!htbp]
\tablecaption{Experimental papers with either no data or with data not included in
the present MARVEL analysis.
Comments about these sources are given in section~\ref{com2}.
}
\label{tab:EXPnot}
\tablewidth{0pt}
\tablehead{
\colhead{Tag}& \colhead{Reference} & \colhead{Range(cm$^{-1}$)} &  \colhead{Method} &\colhead{Data}  & \colhead{Comments} }
\startdata

%\hline
\multicolumn{2}{l}{Phillips \Astate$-$\Xstate}\\
82ErLaMa& \citet{82ErLaMa.C2} &&&No& (3a)\\
\\
\multicolumn{2}{l}{Swan    \dstate$-$\astate}\\
 63CaGi& \citet{63CaGi}&&&No& \\
68MeMe& \citet{68MeMe}&&& No& (3b) \\
  68PhDaxx&\citet{68PhDaxx.C2} &&&Yes& (3c)\\
 88UrBaJa&\citet{88UrBaJa.C2}&21,280-25,930&&No\\
 90ChMa &\citet{90ChMa}&&& No\\
 97KaHuEw&\citet{97KaHuEw} &&    & No & (3d)\\
 94CaDo&\citet{94CaDo}     &&    & No & (3e)\\
 99LlEw & \citet{99LlEw}   &&    & Yes& (3f)\\
 05SmPaSc &\citet{05SmPaSc}&& CL & No\\
\\
\multicolumn{2}{l}{Ballik--Ramsay \bstate$-$\astate}\\
63BaRaa &\citet{63BaRaa.C2} &&& Yes & (3g)\\
\\
%The transition below has no name or reported lines.
\Dstate$-$B$'$~$^{1}\Sigma_{\rm g}^+$\\
91BaUrJa&\citet{91BaUrJa}&28,030-28,555& LIF & No\\
\\
\multicolumn{2}{l}{Duck \dstate$-$\cstate}\\
06KoReMoNa&\citet{06KoReMoNa.C2}&&&No\\

%JONATHAN: THE NEXT TWO SOURCES ARE NOT EVEN MENTIONED IN THE TEXT
\\
\multicolumn{2}{l}{Kable--Schmidt $e^{3}\Pi_{g} -$ \cstate} \\
09NaJoPaRe&\citet{09NaJoPaRe.C2}&&&No&(3h)\\
\\
\multicolumn{2}{l}{Herzberg}\\
69HeLaMa&\citet{69HeLaMa}& 69,000-73,000& FD  & Yes& (3i)\\
\\
\multicolumn{2}{l}{Freymark \Estate$-$\Astate} \\
96BlLiSo&\citet{96BlLiSo.C2}&43,510-43,545&LIF& No.\\
\\
\multicolumn{2}{l}{Deslandres--d'Azambuja \Cstate$-$\Astate}\\
85AnBoPe&\citet{85AnBoPe.C2}&23,800-31,250& E    &     & (3j)        \\
87VaHe& \citet{87VaHe}& 21,000-24,500     & LIF  & No  & (3k)              \\
88UrBaJa&\citet{88UrBaJa.C2}&21,280-25,930& LIF  & No  & (3l) \\
\\
\multicolumn{2}{l}{Messerle--Krauss C$^{'}$~$^{1}\Pi_{\rm g} -$ \Astate}\\
67MeKr&\citet{67MeKrxx.C2}&&&No&(3m)\\
\\
\multicolumn{2}{l}{Goodwin--Cool}\\
88GoCoa& \citet{88GoCoa} &&&No&(3n)\\
88GoCob& \citet{88GoCob} &&& Yes&(3o)\\
89GoCo& \citet{89GoCo} &&& Yes& (3p)\\
\hline

\enddata

\tablecomments{CL = chemiluminescence, E = emission, FD = flash discharge,
LIF = laser-induced fluorescence.}
\end{deluxetable}
\renewcommand{\baselinestretch}{1.25}

\subsection{Comments on the data sources}\label{com2}
First, we make comments on those observed bands of C$_2$ which were not used
in the MARVEL analysis.

The situation with the Deslandres--d'Azambuja band is a remarkable one.
The band was originally
observed more than a century ago \citep{1905DeDa.C2} and, as discussed
in the Introduction, has been observed in a number of astronomical objects.
However, while a high-resolution line list
is available for $^{13}$C$_2$ \citep{85AnBoPe.C2} and there are a number of papers reporting
laboratory observation of the band for $^{12}$C$_2$ \citep{40HeSu,49HoHe,69CiDaRy,88UrBaJa.C2,97SoBlLiXu},
there are no high resolution line data for the band.
A high resolution re-measurement of the band for
$^{12}$C$_2$ would appear to be welcome.

Observation of the Messerle--Krauss band is reported in a single,
short paper \citep{67MeKrxx.C2}, which provides no line data.
Similarly, there are so far no published lines for the
recently detected Kable--Schmidt band \citep{09NaJoPaRe.C2}.

The situation with the three Herzberg and two Goodwin--Cool  VUV bands
is somewhat different.
There are papers reporting detailed spectra for each of these bands \citep{69HeLaMa,88GoCob,89GoCo}.
However, each of them comes from a single uncorroborated measurement.
As these bands probe
upper states which are too high in energy for reliable, independent theoretical predictions, it
was decided to leave their inclusion in a MARVEL analysis for future work.
We note that omitting the energy levels associated with these bands is
not critical for the partition sums and thermodynamic data determined as
part of this study.
% since there are lower-lying electronic states, such as those probed by the Deslandres--d'Azambuja
%and Messerle--Krauss bands, which are also omitted from the present analysis.

We also note that papers by \citet{97SoBlLiXu,07JoNaRe.C2} refer to the
so-called LeBlanc band comprising weak \Dstate $-$ B$'$~$^{1}\Sigma_{\rm g}^{+}$
transitions; however,
we could find no line data on this band or any papers by LeBlanc reporting it.

Second, the set of comments below refer to data sources used in our MARVEL analysis
and listed in Table~2.

\noindent
(2a) 06PeSi \citep{06PeSi}
Uncertainty assumed to be 0.01 \cm\  (Petrova, private communication, 2015).
\\
(2b) 04ChYeWoLi \citep{04ChYeWoLi}
Not the original data are used but those data extracted from 15ChKaBeTa \citep{15ChKaBeTa.C2}.
\\
(2c) 97SoBlLiXu \citep{97SoBlLiXu} and 39Landsver \citep{39Landsver.C2}
Uncertainty of these two sources is assumed to be 1.0 and 0.2 \cm, respectively.
\\
(2d) 51Freymark \citep{51Freymark.C2}
The stated uncertainty was doubled to 0.02 \cm, as this seems to be a more adequate
guess of the accuracy of these transitions.
\\
(2e) 85YaCuMeCa \citep{85YaCuMeCa} Results recorded using magnetic rotation.
The data are provided by Curl (private communication, 2015).
The uncertainty was increased to 0.005 cm$^{-1}$ as values below this did not
give consistent results during the MARVEL analysis.
\\
(2f) 85RoWaMiVe \citep{85RoWaMiVe.C2}
This source presents an analysis of measurements by \citet{79AmChMa.C2};
the data were extracted from 15ChKaBeTa \citep{15ChKaBeTa.C2}.
\\
(2g) 13BoSyKnGe \citep{13BoSyKnGe} New assignments and reassignments of
the measurements by 13YeChWa \citep{13YeChWa.C2}.
\\
(2h) 13YeChWa \citep{13YeChWa.C2} Assignments corrected following
13BoSyKnGe \citep{13BoSyKnGe}.
\\
(2i) 03KaYaGuYu \citep{03KaYaGuYu}    0.007 cm$^{-1}$ uncertainty; includes
 26 transitions with $\Delta \Omega =1$.
\\
(2j)  85CuSa  \citep{85CuSa},
85SuSaHi   \citep{85SuSaHi.C2},
 94PrBe     \citep{94PrBe.C2} and
 10BoKnGe   \citep{10BoKnGe.C2}
Data extracted from 13BrBeScBa \citep{13BrBeScBa.C2}.
%(2j) 11BoSyKnGe \citep{11BoSyKnGe.C2} 23 lines extracted from 13BrBeScBa \citep{13BrBeScBa.C2}.
%23 neglected. \red{The table claims we validated all 46 lines! Comments?}
\\
(2k) 86HaWi \citep{86HaWi}. Data provided by Hardwick (private communication, 2015).
\\
(2l) 98BrHaKoCr \citep{98BrHaKoCr} Uncertainty assumed to be 0.02 \cm.
\\
(2m) 16KrBaWeNa \citet{16KrBaWeNa} give a single line position as part
of their measurement of their multiphoton spectroscopic measurement of the
ionization energy of C$_2$.\\

%\red{According to 16ChKaBeTa \citep{16ChKaBeTa.C2} they study both Bernath bands but we have
%data assigned to only one. Is there something missing?}

\vskip 0.2cm

Third, the next set of comments refer to sources listed
in Table~3; these were not used in our MARVEL analysis for reasons listed below.

\noindent (3a) 82ErLaMa \citep{82ErLaMa.C2} This source contains lifetime data but no
transition frequencies.
\\
(3b) 68MeMe \citep{68MeMe} No actual line data given in the paper
but 11BoSyKnGe \citep{11BoSyKnGe.C2} presents some lines reassigned from this work
which are included in our dataset, see Table~\ref{tab:EXP}.
\\
(3c)  68PhDaxx \citep{68PhDaxx.C2} A book with an extensive list of lines
including higher bands but only 3607 of the 10910 lines could be be validated;
\citet{02TaAm.C2} also found that these assignments did not match those of other work.
It was therefore decided to omit these lines from the final compilation.
\\
(3d)  97KaHuEw \citep{97KaHuEw}
A precursor to 99LlEw  \citep{99LlEw}.
\\
(3e) 94CaDo \citep{94CaDo} So-called high pressure (HP) band.
\\
(3f)  99LlEw  \citep{99LlEw}
None of these data were selected by 13BrBeScBa \citep{13BrBeScBa.C2},
so they were not considered in this study either.
\\
(3g) 63BaRaa \citep{63BaRaa.C2}
Original observation by Ballik and Ramsay of their eponymous band.
The work was re-assigned by \citet{75Veseth}, but his results are not available.
\\
%JONATHAN: THE NEXT TWO SOURCES ARE NOT EVEN MENTIONED IN THE TEXT
(3h) 09NaJoPaRe \citep{09NaJoPaRe.C2}
Report of a new C$_2$ band but no data are provided and no follow-up study exists.
\\
(3i) 69HeLaMa \citep{69HeLaMa}
Report of three new VUV bands with line data.
Upper states lie at too high energy for data to be validated.
\\
(3j) 85AnBoPe \citep{85AnBoPe.C2} Report of extensive data for $^{13}$C$_2$ but with no
dataset for $^{12}$C$_2$.
\\
(3k) 87VaHe \citep{87VaHe} Report of a \lq\lq new\rq\rq\ C$_2$ band:
25 transitions from one (unknown) band were recorded
with an uncertainty of 0.2 \cm\    and assigned $J$ quantum numbers; these were
supplied by Heaven (private communication, 2015). The calculations of \citet{92BrWr.C2} suggest that
this band actually belongs to C$_2^+$ rather than C$_2$.
The band remains unassigned and the data were not included in our current analysis.
\\
(3l) 88UrBaJa \citep{88UrBaJa.C2}  No transition data reported in the paper.
\\
(3m) 67MeKr \citep{67MeKrxx.C2}
Discovery paper giving many spectroscopic parameters but no primary transition data.
\\
(3n) 88GoCoa \citep{88GoCoa} Discovery paper:  follow-up work with data in
88GoCob \citep{88GoCob} and 89GoCo \citep{89GoCo}.
\\
(3o) 88GoCob \citep{88GoCob} New band 1 $^1\Delta_{\rm u}-$ A $^1\Pi_{\rm u}$ with line data.
Upper states lie at too high energy for data to be validated.
\\
(3p) 89GoCo \citep{89GoCo} New band 1 $^1\Delta_{\rm u}-$ B $^1\Delta_{\rm g}$ with line data.
Upper states lie at too high energy for data to be validated.

\subsection{Rovibronic nuclear motion computations using \Duo}
In order to decide on their correctness,
we have compared the experimental MARVEL rovibronic energies with their theoretical
counterparts.
The latter approximate but complete set of energy levels is based on
empirical potential energy curves (PEC), spin-orbit curves (SOC),
and electronic angular momentum curves (EAMC) of C$_2$
as given by \citet{jtexoC2}.
The theoretical rovibronic energies were computed using a new diatomic nuclear motion
program called \Duo\ \citep{jt609}.
\Duo\ solves the fully-coupled rovibronic Schr\"{o}dinger equation variationally
using a combination of discrete variable representation (DVR) and
rigid-rotor basis sets to represent the vibrational and the
spin-rotational degrees of freedom in the Hund's case~(a) representation, respectively.
The final PECs, SOCs, and EAMCs were obtained by refining \ai\ curves
obtained at the ic-MRCI/aug-cc-pVQZ level of electronic structure theory
for the nine lowest electronic states of C$_2$,
\Xstate, \Astate, \Bstate, B$'$~$^{1}\Sigma_{\rm g}^+$, \astate, \bstate, \cstate, \dstate, and $1~^{5}\Pi_{\rm g} -$ by fitting to the MARVEL energies.
A detailed account of these computations will be reported elsewhere.

\Duo\ uses the following quantum numbers motivated by the Hund's case (a)
choice of the basis set:
$$
\{ J, \, +/-, \, {\rm state}, \, v, \, \Lambda, \, \Sigma, \, \Omega \}
$$
where $\Lambda$, $\Sigma$, and $\Omega (\Omega=\Lambda+\Sigma)$ are the
signed quantum numbers corresponding to projections of the electronic,
spin, and total angular momenta, respectively
(the projection of the rotational angular momentum $\hat{R}$ on the
molecular axis is zero).
The \Duo\ quantum numbers characterizing the computed rovibronic states
are used to check and complete the MARVEL labels.
It should be noted that \Duo\ uses an approximate assignment
scheme based on the largest contribution to the wavefunction expansion \citep{jt609}.
Although this scheme is very robust,  it can sometimes lead to ambiguous sets of
quantum  numbers, especially for states in strong resonance with other rovibronic states.

Comparison of the \Duo\    and MARVEL results helped us to identify problems in the
experimental data, such as misassigned lines, duplicate transitions,
and outliers.

\begin{figure}
\includegraphics[height=125mm]{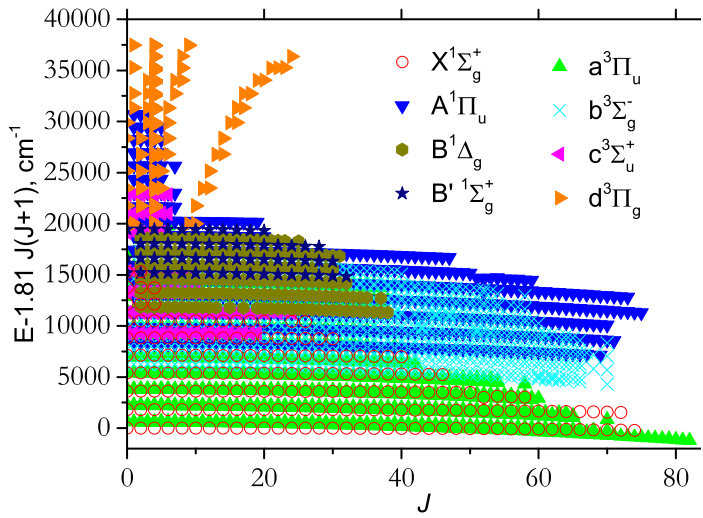}
\caption{The MARVEL term values for the eight lowest-energy
singlet and triplet states   shown
as a reduced-energy diagram, after subtraction of  $1.81 J(J+1)$ \cm\   from
the energy $E$ of the state.}
\label{fig:MarvelELs}
\end{figure}

\renewcommand{\baselinestretch}{1.0}
\begin{deluxetable}{lrrr|crrr}%[!htbp]
\tablecaption{Empirical, experimental (MARVEL), and theoretical $T_{\rm e}$ and $T_{0}$ term
values, in \cm, of the six lowest-energy singlet and triplet and the
two lowest-energy quintet electronic states of C$_2$.}
\label{tab:Tes}
%\tablewidth{0pt}
\tablehead{
\colhead{State}    &  \colhead{}    & \colhead{$T_{\rm e}$} &  \colhead{}   &\multicolumn{2}{c}{$T_{0}$}} %\\
\startdata
&Calc.$^{a}$& Empirical & \Duo$^c$&Empirical & MARVEL$^{c,d}$& $J$  & $F$\\
\Xstate            &  0.0           & 0.0                   &     0.0       &    0.0       				& 0.0         & 0    & 1\\
\astate            & 509            & 720.0083(21)$^{e}$   	&    722.58     &  603.828$^{b}$      & 603.817(4) 	 & 2    & 1\\
\bstate            & 6233           & 6439.08382(58)$^{e}$  &   6437.00     &  6250.164$^{b}$     & 6250.149(7)  & 0    & 3\\
\Astate            & 8374           & 8391.4062(19)$^{e}$   &   8393.63     &  8271.606$^{b}$     & 8271.607(7)   & 1    & 1\\
\cstate            & 9371           & 9124.2$^{f}$    			&   9172.30     &  9277$^{j}$        & 9280.215(5)    & 1    & 2\\
\Bstate            &11966           & 12082.34355(54)$^{g}$ &  12092.45     & 11859$^{j}$        & 11867.825(5)  & 2    & 1\\
B$'$~$^{1}\Sigma_{\rm g}^+$&15261   & 15410.33(36)$^{g}$   	&  15401.52     & 15197$^{j}$        & 15196.509(5) & 0    & 1\\
d~$^{3}\Pi_{\rm g}$ &20092          & 20022.5$^{f}$   			&  20030.92     & 19992$^{j}$        & 19983.953(8) & 2    & 1\\
$1~^{5}\Pi_{\rm g}$ &               &   &               & 29258.5922(48)$^{h}$& 29860.921(5) &2    & 1\\
\estate             &               & 40796.7$^{f}$         &               &              & 40422.691 (50)   & 2    & 1\\
\Dstate             &               & 43239.8$^{f}$         &               &              & 43230.499 (138)  & 1    & 1\\
$1~^{5}\Pi_{\rm u}$ &               & &               &  51 049.799$^{i}$	& 51651.142(20) & 1    & 1\\
$4~^{3}\Pi_{\rm g}$ &               &              					&               &              & 52106.042(94) & 1    & 1\\
%D$~^{1}\Delta_{\rm g}$&             &              &    \\
E$~^{1}\Sigma_{\rm g}^{+}$&         & 55034.7$^{f}$					&               &              & 54936.672(150) & 0    & 1\\
%$F~^{1}\Pi_{\rm u}$ &               &              &               &               \\
\enddata

\tablecomments{$^{a}$ \citet{01MuDaLiDu}, MR-CISD values at the complete basis set (CBS) limit.
$^{b}$ Vibronic energies taken from the Suppl. Mat. of \citet{15ChKaBeTa.C2}.
$^{c}$ This work.
$^{d}$ The lowest observed energy level in each electronic state is reported.
%{\it e.g.}, for v=0, J=1, e parity for the \Astate. Is this true?
$^{e}$ From \citet{15ChKaBeTa.C2}.
$^{f}$ From \citet{09BaRiPe.partfunc}.
$^{g}$ From Chen et al. (2016).
$^{h}$ From \citet{11BoSyKnGe.C2}, relative to the $v=0$ level of the \astate\    state.
$^{i}$ From \citet{15BoMaGo}, relative to the $v=0$ level of the \astate\   state.
$^{j}$ Empirical vibronic energies taken from \citet{15ChKaBeTa.C2} }
\end{deluxetable}

\section{Results and discussion}

\subsection{Term values}
In its ground electronic state C$_2$ has a fairly strong
bond \citep{39Mulliken.C2,01MuDaLiDu,05ShPi,11SuWuGuWu}.
The equilibrium bond distance of the \Xstate\   state at 1.2425 \AA\   is short
for a double bond (considerably shorter than the C=C bond of
ethylene at $r_{\rm e}$(C=C) = 1.331 \AA),
but long for a triple bond (in C$_2$H$_2$ $r_{\rm e}$(CC) = 1.203 \AA).
This also means that the dissociation energy of the \Xstate\   state is large,
in fact more than 50,000 \cm.
Compared to this, the energy differences among the several feasible
asymptotes comprised by the low-lying $^3$P, $^1$D, and $^1$S states
of the C atom are relatively minor.
Furthermore, the structure of the molecular orbitals (MO) of C$_2$ is such
that a large number of low-energy singlet, triplet, and quintet valence states
are feasible and many of them are part of experimentally measurable
rovibronic transitions (see Table~1).
In the \Xstate\  state the leading valence electron configuration is
(core)$(2\sigma_{\rm g})^2(2\sigma_{\rm u})^2(1\pi_{\rm u})^4$.
%(3\sigma_g)^1$
By promoting electrons from the weakly antibonding $2\sigma_{\rm u}$ and the strongly
bonding $1\pi_{\rm u}$ MOs and populating the
%antibonding
$3\sigma_{\rm g}$ MO a large number of electronic states arise.
Fortunately, for obtaining proper, temperature-dependent ideal-gas thermochemical
quantities up to about 4,000 K, it is sufficient to consider 9
electronic states, four singlet, four triplet, and one quintet states (see Table 4).
As mentioned already,
all these lowest-energy electronic states correlate with the C($^3$P) + C($^3$P)
separated-atom limit.

Not too surprisingly for such a simple molecule, a large number of
electronic structure computations are available for C$_2$ in the
literature \citep{79KiLi.C2,87BaLa.C2,92WaBart.C2,00BoVoHa.C2,01BrGrein.C2,01MuDaLiDu,04AbSh.C2,05ShPixx.C2,11JiWi,13BoThRu.C2,14BoThRuWi}.
There are several issues which make the electronic structure computations
extremely challenging for C$_2$.
First,
there is a quasi-degeneracy of the fully occupied $1\pi_{u}$ and
the empty $3\sigma_{g}$ MOs,
explaining some of the unusual characteristics of the excited electronic states of C$_2$.
Second,
the existence of several low-lying excited electronic states leads to
the occurrence of a considerable number of avoided crossings
among the PECs as the CC distance is varied.
Third,
as pointed out by \citet{04AbSh.C2} based on full configuration interaction (FCI) computations,
at least in the cases of the X, B, and B$'$ states,
methods based on an unrestricted Hartree--Fock (UHF)
reference provide correct but methods based on a restricted HF (RHF)
reference provide incorrect results.
Fourth,
one must account for the strong multireference character of
the electronic states and the near degeneracies changing rapidly along
the CC distance.
Fifth,
rather large atom-centered, fixed-exponent Gaussian basis sets are
required for the correct and converged description of the valence states.
These difficulties explain why this deceptively simple diatomic molecule
is still one of the favorites of developers of modern wavefunction-based
electronic structure techniques \citep{92WaBart.C2,01MuDaLiDu,14BoThRuWi}.

Using CI methods, 27 bound valence states of C$_2$
were computed by \citet{66FoNe}.
\citet{79KiLi.C2} obtained results for all 62 electronic states
in the valence manifold,  including weakly bound and repulsive ones.
\citet{83PoRoScRo}  obtained results for Rydberg states, as well.
Electronic states with $T_{\rm e}$ values up to 75,000 \cm\ (this is in fact the
F~$^1\Pi_{\rm u}$ state) have been studied but for the present investigation
the energy cut-off value was chosen to be 35,000 \cm.
This limits the number of singlet, triplet, and quintet states to 4, 4, and 1,
respectively, 9 states altogether.
None of the higher-lying states will be considered in what follows.
%All the states considered here correspond to the C($^3P$) + C($^3P$) asymptote.
Note that
RKR potential curves are given for several singlet, triplet, and quintet
states in \citet{92Martin}.

In this study, the so-called $T_0$ values obtained for the electronic states
define directly the lowest measurable term energies of the states (thus,
they may not necessarily correspond to $J=0$).
It is not that simple to determine the $T_{\rm e}$ values of the excited
electronic states of C$_2$, as these are not measurable quantities.
This can only be achieved if the zero-point vibrational energy (ZPVE) of all
the states is determined.
However, since all states are coupled in the \Duo\ computations,
these do not come directly from our joint MARVEL and \Duo\ analysis.
In particular,
for the singlet $\Pi$ and $\Delta$ electronic states there are no transitions
to $J=0$ upper rovibronic states.
% so the lowest observed energy levels are reported in Table 4.

%\newpage
\renewcommand{\baselinestretch}{1.0}
\begin{deluxetable}{llrrrclcl}%[!htbp]
\tablecaption{Vibrational energy levels of the
%five lowest-energy
X~$^{1}\Sigma_{\rm g}^{+}$, B$'~^{1}\Sigma_{g}^+$, and E~$^{1}\Sigma_{g}^{+}$
states of C$_2$.
All values are given in \cm.
All energies are relative to the appropriate $v=0$ vibrational level.}
\label{tab:vib_singletFirst}
\tablewidth{0pt}
\tablehead{
\colhead{State}    & $v$ &\colhead{Calc.$^a$}&\colhead{Expt.$^b$}&  \colhead{MARVEL$^c$}}
\startdata
X~$^{1}\Sigma_{\rm g}^{+}$& 1&  1829.15     & 1827.4849(2)   &  1827.486(5) \\
                       & 2   &  3630.35     & 3626.6835(2)   &  3626.681(10) \\
                       & 3   &  5402.78     & 5396.6892(4)   &  5396.686(9) \\
                       & 4   &  7145.40     & 7136.3507(6)   &  7136.350(6) \\
                       & 5   &  8856.84     & 8844.1241(11)  &  8844.124(7) \\
                       & 6   & 10536.37     & 10517.9659(39) & 10517.950(7) \\
                       & 7   & 12178.69     & 12154.9615(29) & 12154.961(6) \\
                       & 8   & 13783.76     & 13751.3944(38) & 13751.393(3) \\
                       & 9   & 15346.69     & 15302.8952(46) & 15302.893(7) \\
B$'~^{1}\Sigma_{\rm g}^+$  & 1   &              & 1420.4850(4)   & 1420.488(9) \\
											 & 2   &              & 2840.0048(4)   & \\
                       & 3   &              & 4261.0686(4)   & 4261.071(1) \\
											 & 4   &              & 5681.5113(6)   & \\
E~$^{1}\Sigma_{\rm g}^{+}$ & 1   &              &              & 1592.316(200) \\
\enddata

\tablecomments{
$^a$ From 07KoBaSc \citep{07KoBaSc.C2}.
$^b$ From \citet{16ChKaBeTa.C2}.
$^c$ This work,  values of the $v=0$ vibrational levels of X~$^{1}\Sigma_{\rm g}^{+}$, B$'~^{1}\Sigma_{\rm g}^+$,
and E~$^{1}\Sigma_{g}^{+}$ are, in \cm, 0.0, 15196.509, and 54936.664, respectively.}
\end{deluxetable}

%\newpage
\renewcommand{\baselinestretch}{1.0}
\begin{deluxetable}{lrrrllrrr}%[!htbp]
\tablecaption{The lowest-energy states with $J \neq 0$ of the excited vibrational levels
of the A~$^{1}\Pi_{\rm u}$ and B~$^{1}\Delta_{\rm g}$ singlet states of C$_2$.
All values are given in \cm. All energies are relative to the appropriate $v=0$ vibrational level.}
\label{tab:vib_singletSecond}
\tablewidth{0pt}
\tablehead{
\colhead{State}        & $v$& $J$ &  \colhead{MARVEL$^a$}&& State      &$v$ &  $J$ & \colhead{MARVEL$^a$} }
\startdata
A~$^{1}\Pi_{\rm u}$    & 1  & 1    & 1584.008(4)   &&B~$^{1}\Delta_{g}$& 1  & 2    & 1384.440(2)  \\
                       & 2  & 1    & 3143.805(5)   &&                  & 2  & 2    & 2746.010(9)  \\
                       & 3  & 1    & 4679.323(6)   &&                  & 3  & 2    & 2746.010(9)  \\
                       & 4  & 1    & 6190.503(7)   &&                  & 4  & 2    & 5400.804(8)  \\
											 & 5  & 1    & 7677.273(4)   &&                  & 5  & 2    & 6694.148(10) \\
											 & 6  & 1    & 9139.523(9)   &&                  & 6  & 2    & 7964.836(3)  \\
											 & 7  & 1    & 10577.184(13) &&                  & 7  & 2    & 9214.047(3)  \\
											 & 8  &      &               &&                  & 8  & 3    & 10446.668(5) \\
											 & 9  & 1    & 13378.377(1)  &&                  & \\
											 & 10 & 1    & 14741.688(1)  &&                  & \\
											 & 11 & 1    & 16079.978(7)  &&                  &\\
											 & 12 & 1    & 17393.116(8)  &&                  & \\
											 & 13 & 1    & 18680.944(7)  &&                  & \\
											 & 14 & 1    & 19943.297(7)  &&                  & \\
											 & 15 & 1    & 21179.968(7)  &&                  & \\
											 & 16 & 1    & 22390.754(7)  &&                  & \\
%     & 1  & 2    & 1384.440(2)  \\
%                       & 2  & 2    & 2746.010(9)  \\
%											 & 3  & 2    & 2746.010(9)  \\
%											 & 4  & 2    & 5400.804(8)  \\
%											 & 5  & 2    & 6694.148(10)  \\
%											 & 6  & 2    & 7964.836(3)  \\
%											 & 7  & 2    & 9214.047(3)  \\
%											 & 8  & 3    & 10446.668(5)  \\
%D~$^{1}\Sigma_{u}^+$    & ??   &  ??            &      ??        & \\
\enddata

\tablecomments{$^a$ $J$ is the quantum number corresponding to the total angular momentum.
The values of the $v=0$ vibrational levels of A~$^{1}\Pi_{\rm u}$ and B~$^{1}\Delta_{g}$
are 8271.607(7) \cm\  ($J=1$) and 11867.825(5) \cm\  ($J=2$), respectively.
The rule $J \geq |\Omega|$ for Hund's case (a) coupling explains the lack of the $J=0$ states for
\Astate\   and the $J=0,1$ states for \Bstate.}
\end{deluxetable}

%\newpage
\renewcommand{\baselinestretch}{1.0}
\begin{deluxetable}{lrrr}%[!htbp]
\tablecaption{Excited vibronic levels for the four lowest-energy
triplet states of C$_2$ with $J=0$ and $F_3$.
All values are given in \cm.
Energies in each electronic state are relative to the appropriate $v=0$ ($J=0,F_3$) vibrational level.}
\label{tab:vib_triplet}
\tablewidth{0pt}
\tablehead{
\colhead{State}    & $v$ &  \colhead{MARVEL$^a$}}
\startdata
a~$^{3}\Pi_{\rm u}$    & 1   &  1617.985(10) \\
                       & 2   &  3212.620(9) \\
											 & 3   &  4783.940(5) \\
											 & 4   &  6331.973(1) \\
											 & 5   &  7855.893(100) \\
											 & 7   & 10835.820(1) \\
											 & 8   & 12290.393(1) \\
											 & 9   & 13721.623(5) \\
											 &10   & 15129.564(1) \\
											 &11   & 16514.369(1) \\
b~$^{3}\Sigma_{\rm g}^{-}$& 1&  1448.103(8)  \\
                       & 2   &  2874.028(2) \\
                       & 3   &  4277.927(1) \\
c~$^{3}\Sigma_{\rm u}^{+}$& 1&  2031.833(8) \\
                       & 2   &  4034.776(7) \\
											 & 3   &  6007.745(8) \\
											 & 5   &  9859.060(1) \\
											 & 6   & 11734.338(10) \\
											 & 7   & 13573.601(1) \\
d~$^{3}\Pi_{\rm g}$    & 1   &  1753.500(7)  \\
                       & 2   &  3469.636(10) \\
											 & 3   &  5145.247(6) \\
											 & 4   &  6776.153(1) \\
											 & 5   &  8356.139(7) \\
											 & 6   &  9880.362(7) \\
											 & 7   & 11337.658(1) \\
											 & 8   & 12722.024(1) \\
											 & 9   & 14025.567(1) \\
											 &10   & 15245.388(2) \\
%$e^{3}\Pi_{g}$                  &  589              &  366                          & 223 & \\
\enddata

\tablecomments{
$^a$
%Each energy belongs to the $F = 3$ spin component for $^3\Pi$ states with J=? and ? parity. For $^3\Sigma$ states the $F=1$, N=0, J=1 level is used. (This needs to be checked.)
The ($v=0,J=0,F_3$) energy values are 632.730, 6250.149, 9280.834, and 20009.011 for the a~$^{3}\Pi_{\rm u}$,
b~$^{3}\Sigma_{\rm g}^{-}$, c~$^{3}\Sigma_{\rm u}^{+}$, and d~$^{3}\Pi_{\rm g}$ states, respectively. }
\end{deluxetable}

%\newpage
%\renewcommand{\baselinestretch}{1.0}
%\begin{deluxetable}{llrrrclcl}%[!htbp]
%\tablecaption{Vibrational energy levels of the two lowest-energy quintet states of %C$_2$.}\label{tab:vib0_triplet}
%\tablewidth{0pt}
%\tablehead{
%\colhead{El. state}    & $v$ &\colhead{Calc.$^a$}&\colhead{Expt.$^b$}&  \colhead{MARVEL}}
%\startdata
%1~$^{5}\Pi_{\rm g}$    & 1   &       &       &   \\
%                       & 2   &   \\
%1~$^{5}\Pi_{\rm u}$    & 1&       &       &    \\
%                       & 2   &       &       &   \\
%\enddata
%
%\tablecomments{$^a$ From 07KoBaSc \cite{07KoBaSc.C2}. $^b$ From 07KoBaSc \cite{07KoBaSc.C2}.}
%\end{deluxetable}

\subsection{Vibrational energy levels}

Figure~\ref{fig:MarvelELs} shows all MARVEL term values below 35,000~\cm\  for
four singlet and four triplet states,
where $1.81 J(J+1)$ cm$^{-1}$ has been subtracted from the computed energies to make
the figure clearer: this means that near-horizontal sequences
of levels for a particular electronic state are all associated with
something one could call a single vibrational level.
%\red{I don't understand what is supposed to be below 35~000~cm$^{-1}$. It is
%for the 8 lowest states but there are (a) further states below 35~000~cm$^{-1}$
%and (b) there levels on the figure above 35~000~cm$^{-1}$.}
Figure~\ref{fig:MarvelELs} shows that the largest total angular momentum quantum
numbers, $J_{\rm max}$, are 74, 75, 86, and 70 for the \Xstate, \Astate,
\astate, and \bstate\   states, respectively.
As expected, as the vibrational excitation increases, the
$J_{\rm  max}$ value usually decreases.
Finally, note that the coverage of rovibronic levels up to 35,000 \cm\ from
experiment is not complete;
assuming rigid rotation, data up to about $J=144$ is needed to have
full coverage of the energy levels required during the thermochemical analysis.
This coverage is provided in this study by \Duo\ energy levels ({\it vide infra}).

%\red{Tables 5 and 6, I would comment:
%(a) why do we not give data for electronic states studied? (b) do we use vibrational
%values relative to a stated $T_0$? If so we must say so, (c) we need a Table 7 for
%quintets.}

Due to the strength of the CC bond in all the electronic states studied,
the vibrational fundamental is substantial in almost all the bound electronic
states of C$_2$.
In fact, for the ground electronic state the harmonic wavenumber is close to
2,000 \cm, a high value for a relatively heavy molecule.
Thus, the number of vibrational states is not that high, despite the large
dissociation energy.
In particular,  \citet{14BoThRuWi} computed  57, 54, 49, and 36
bound vibrational levels for the \Xstate, \Astate, \Bstate, and
B$'$~$^{1}\Sigma_{\rm g}^+$ states, respectively.
The number of vibrational levels characterized by our MARVEL analysis is
considerably smaller, only 9, 16, 8, and 3, respectively.
The vibrational energies presented for the triplet electronic states given in Table~7
%\ref{tab:vib_triplet}
can not be compared easily with existing literature values, since the MARVEL values
are for a specific spin component of a rovibronic energy level.

The largest ``vibrational fundamental'' ($J=0$) corresponding to the electronic states studied
here is that of the c~$^{3}\Sigma_{\rm u}^{+}$ state, at 2031.833~\cm.
Consequently, this state must have the strongest CC bond.

%\red{SY: How did we derive the triplet `vibrational energies'? They are not actually they cannot be experimental
%or MARVEL, since these energies do not seem to have the spin-components. Do I miss something? Are they band-centres?
%I think we should explain this also how they have been estimated.}
%It can be seen that the MARVEL vibrational energies are significantly different
%from empirically determined ones of \citet{13BrBeScBa.C2}
%\red{PFB: need to compare with Nakajima and Endo}
%(they report levels up to $v=9$ only while MARVEL
%energies are available up to $v=11$), and that
%the difference grows with the vibrational quantum number $v$.
%\red{Peter disagrees with the following sentence and suggests it is
%deleted}
%This suggests that the molecular constants of \citet{13BrBeScBa.C2}
%are results of a fitting procedure with unreasonably small
%uncertainties and do not reflect accurately the true vibrational
%energies of the \astate\ state of C$_2$.

%\newpage
\begin{figure}
\includegraphics[height=125mm]{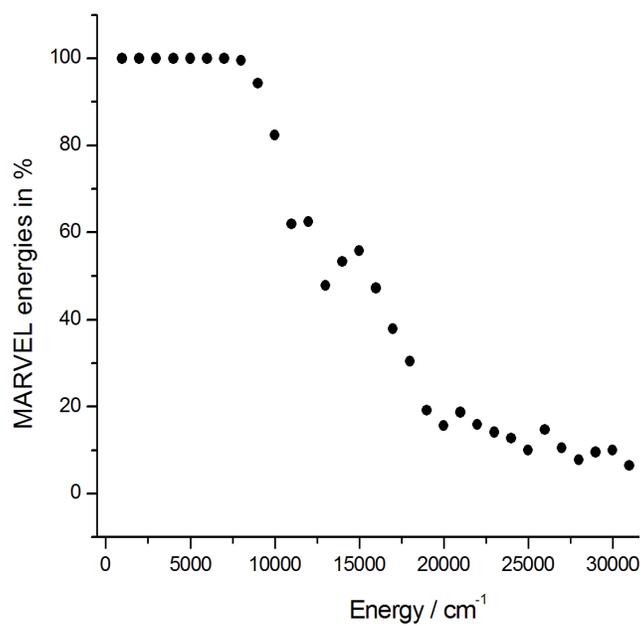}
\caption{Completeness of the rovibronic MARVEL energies as a function of the excitation energy.}
\label{fig:MarvelRovib}
\end{figure}

\subsection{Rovibrational energy levels}

The rotational constant of C$_2$ for the \Xstate\ state is relatively small,
about 1.81 \cm.
This results in a large number of rotational states for each vibrational level.
As part of this study, \Duo\ results were obtained up to $J=144$.
They served to check whether a MARVEL energy level is viable or not as well as
they were generated to help the thermochemical analysis of this study ({\it vide infra}).

The lowest missing MARVEL energy level is at 7,800 \cm;
up to this energy the coverage is complete.
As Figure~\ref{fig:MarvelRovib} shows,
as the energy increases there are more and more experimentally unknown energy levels.
The coverage drops below 10\% at about 30,000 \cm.

It is interesting to note how close some of the rovibrational energy differences
are to each other.
Take the \astate\  state as an example.
The $v=1 - v=0$ energy difference for $J=0$ is 1617.985 \cm.
The highly similar energy differences for the different spin components
($J=1,F=2$), ($J=1,F=3$), ($J=2,F=1$), ($J=2,F=2$), and ($J=2,F=3$), are
1618.063, 1617.941, 1617.985, 1617.902, and 1617.831, respectively.
The reason behind this observation is that each spin component of each
vibrational level of
the \astate\  state has a slightly different effective
value for their rotational constant due to both
electronic and vibrational effects, and these energy differences include
rotational energies.
The first-principles \Duo\   energies, where such interactions are taken
explicitly into account, results in similar energy differences.

\section{Thermochemical properties of C$_2$}

\citet{60Altman} and \citet{61Clementi} seem to be the first
to address the thermochemical properties of the C$_2$ molecule.
They both based their analysis on spectroscopic constants available to them
and included several, but not all necessary, electronic states in their study.
More recent thermochemical studies include those
by \citet{81Irwin.partfunc},  \citet{84SaTaxx.partfunc}, and \citet{85RoMaVe.partfunc}.
The most reliable results appear to have been given by \citet{90Gurvich}.

The internal partition function, $Q_{\rm int}(T)$, of C$_2$ is computed here
via the direct summation recipe \citep{jt263} using a mixture of experimental (MARVEL)
and theoretical (\Duo) energy levels .
\Duo\ provides the full set of energies for the nine electronic states considered
during the determination of the ideal-gas thermochemistry of C$_2$.
While this is a complete set, the energy levels are of limited accuracy.
Thus, whenever possible, the \Duo\ energy levels are replaced by the incomplete but
accurate set of MARVEL rovibronic energies.
The final energies are used to compute the internal partition sum,
its first ($Q'$) and second ($Q''$) moments, and the
specific heat ($C_{\rm p}$) as a function of temperature.

Inaccuracies in $Q_{\rm int}(T)$ have three sources of origin.
The first is the intrinsic uncertainty of the energy levels.
%used to calculate the partition function.
The second is the lack of a complete set of bound rovibronic energy levels.
%up to the dissociation limit for all the electronic states.
The third is associated with the treatment, including the possible neglect,
of the unbound states.

\begin{figure}
\includegraphics[height=125mm]{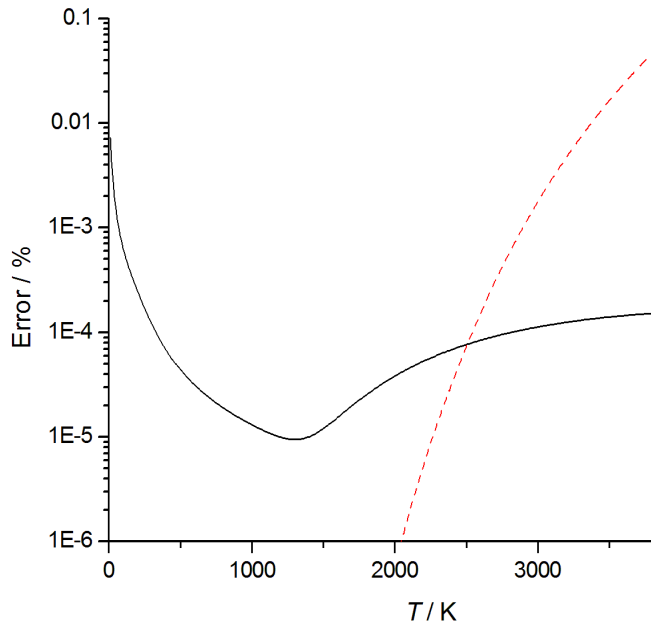}
\caption{Sources of uncertainty in the internal partition function
of $^{12}$C$_2$ up to 4000 K.
The solid (black) line shows the uncertainty which arises from the uncertainties
of the known energy levels.
The dashed (red) curve represents the convergence error, in \%, due to energy
levels not included in the analysis.}
\label{fig:QError}
\end{figure}

The main source of uncertainty in $Q_{\rm int}(T)$ can be estimated straightforwardly
using the uncertainty of each energy level and an error propagation formula.
All experimental (MARVEL) energy levels have an associated uncertainty,
while an uncertainty of 5.0 \cm\   was assumed for the uncertainties of all the DUO levels.
(Note that this uncertainty estimate is rather pessimistic.)
Figure~\ref{fig:QError} (solid line) shows the impact of the uncertainties of
the energy levels on the uncertainty of $Q_{\rm int}(T)$.
It can be seen that up to about 2,500 K this type of uncertainty is dominant,
but its maximum value, occurring at the lowest temperatures, is still less than 0.01\%.
Estimation of the second type of error is hard since
(a) the exact number of the energy levels is unknown, and
(b) the value of the partition function grows monotonically as more and more energy levels
are considered in the direct sum.
Therefore, only an approximate convergence can be reached at higher temperatures
during the direct summation.
To check the convergence of the partition function we need a larger set of energy levels;
therefore, we computed approximate rovibronic energy levels for all the
electronic states considered, using the spectroscopic constants
published by \citet{90Gurvich} up to 70,000 \cm.
Using the spectroscopic constants of 27 electronic states we could determine 332~347 extra
energy levels above 35,000 \cm.
Figure~\ref{fig:QError} (dashed line) shows the difference (in \%) of the two data sets,
(MARVEL + DUO) and (MARVEL + DUO + approximate energy levels).
Although up to 2,000 K the partition function is fully converged
(the difference is less than $10^{-6}$ \%),
above 2,000 K the difference begins growing appreciably.
Nevertheless, the maximum uncertainty is still less than 0.1\% at 4,000 K, which is
acceptable for probably all practical applications.
In case of C$_2$, no consideration of unbound states \citep{15SzCsxx.MgH} is necessary,
due to the large dissociation energy of C$_2$.

The $Q$, $Q'$, $Q''$, and $C_{\rm p}$ results are given in Table 10 in 100 K intervals.
The full set of results at 1 K
increments is given in the machine readable version of the table in the
electronic edition.

\begin{figure}
\includegraphics[height=125mm]{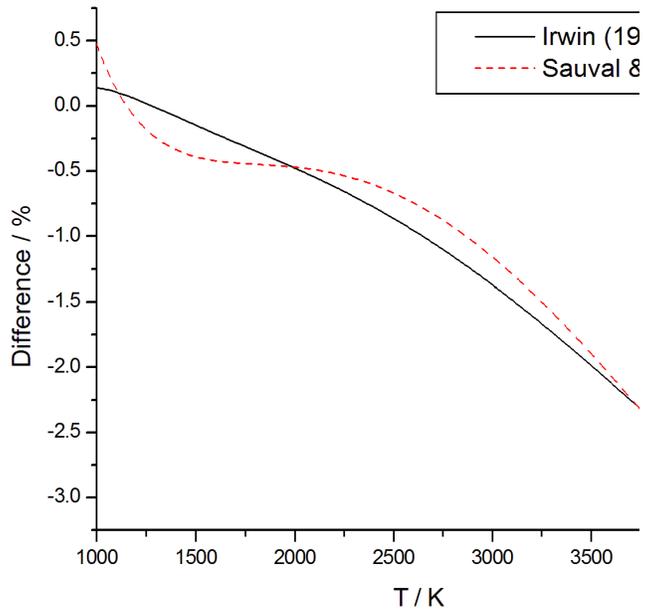}
\caption{The percentage differences of our partition function from \citet{81Irwin.partfunc}
(solid, black) and from \citet{84SaTaxx.partfunc} (dashed, red). }
\label{fig:QDiff}
\end{figure}

Figure~\ref{fig:QDiff} shows the result of the comparison of our $Q_{\rm int}(T)$ with
those of \citet{81Irwin.partfunc} and  \citet{84SaTaxx.partfunc}.
It can be seen that the earlier studies always yield smaller numbers for the
partition function of C$_2$ than the present study.
A possible explanation is that we use a larger (more complete) set of energy levels.
Note that the temperature range of the two earlier works begins from 1,000 K.

\begin{figure}
\includegraphics[height=125mm]{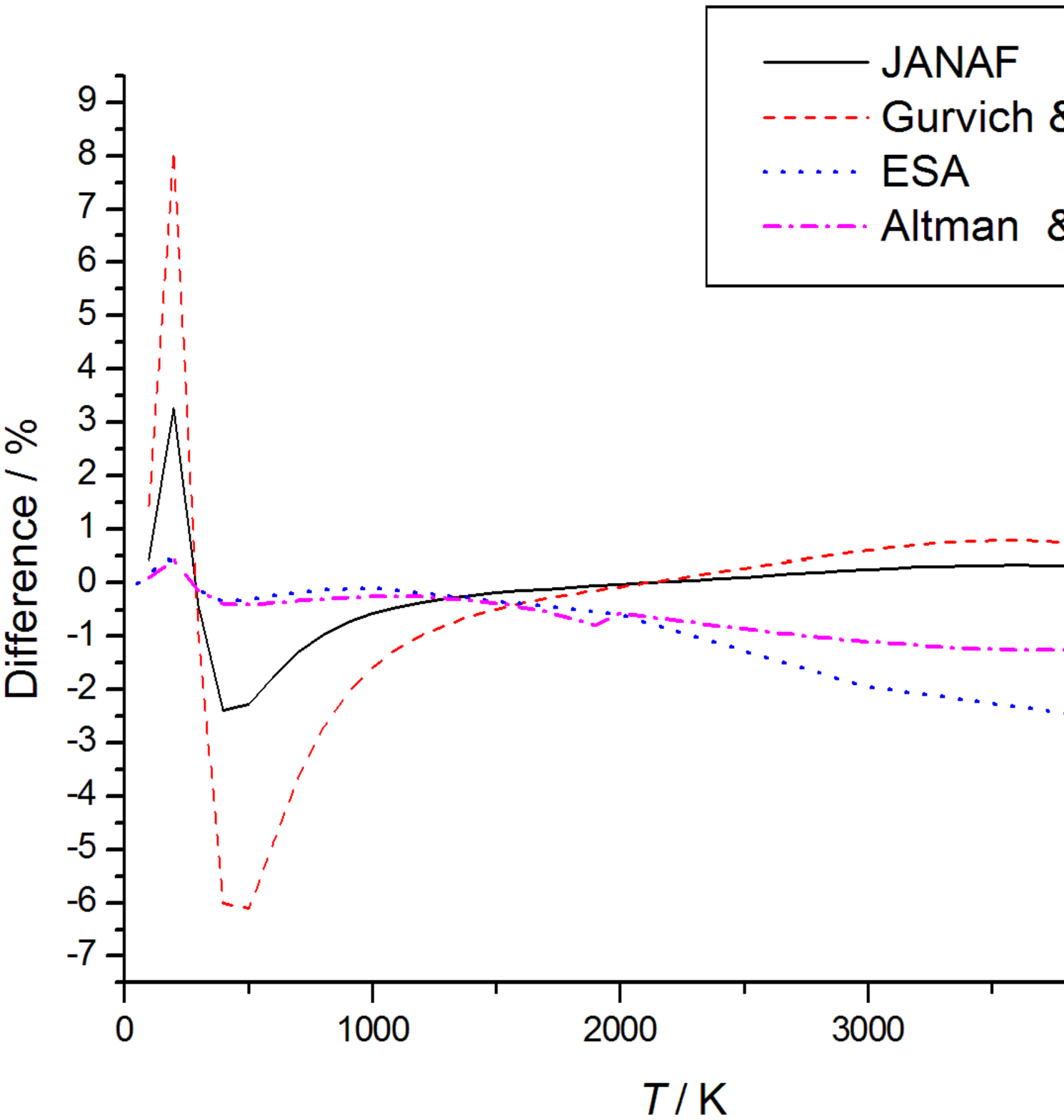}
\caption{Comparison of our $C_{\rm p}$ results with JANAF \citep{janaf} (solid), \citet{90Gurvich} (dashed), ESA \citep{ESA.partfunc} (dotted), and \citet{60Altman} \& \citet{61Clementi} (dashed-dotted, see text) values. }
\label{fig:CpDiff}
\end{figure}

Figure~\ref{fig:CpDiff} shows the difference between our $C_{\rm p}$ results and
the JANAF \citep{janaf}, \citet{90Gurvich}, ESA \citep{ESA.partfunc}, and
\citet{60Altman} and \citet{61Clementi} data.
\citet{60Altman} reported $C_{\rm p}$ values from 0 K to 6,000 K,
but \citet{61Clementi} corrected his values above 2,000 K.
%It is interesting to note that
While there is good agreement with the
ESA and Altman \& Clementi data at low temperatures, the difference begins
to increase with the temperature.
Conversely, for Gurvich and JANAF, the differences,
especially compared to those of Gurvich \& Veyts,
are surprisingly large at lower temperatures.
\citet{09BaRiPe.partfunc} also found this discrepancy
cpncerning the Gurvich \& Veyts (NASA) data at lower temperatures;
nevertheless, they could not explain this strange behavior.
We believe that the problem originates from the incorrect usage of
$T_e$ values in \citet{90Gurvich}.
The standard (spectroscopic) energy expansions use $T_{\rm e}$ as a minimum-to-minimum
excitation energy; in this case the rovibronic energy levels of the upper electronic
state will be shifted by the difference of the zero-point energies,
by about 0.5 $\Delta \omega_{\rm e}$.
To get the correct energies,
the $T_0$ values should be used instead of the $T_{\rm e}$ values.
The first excited electronic state usually lies much above the ground state;
therefore, this relatively small shift does not cause a significant problem
if left out of consideration.
However, in the case of C$_2$, where the relative energy of the \astate\   state
is smaller than the vibrational fundamental of either state,
the incorrect use of $T_{\rm e}$ leads to wrong $C_{\rm p}$ values
at lower temperatures.

Table~8 gives coefficients of the least squares fit
to our computed partition function using the traditional form of \citet{jt263}
\begin{eqnarray}
{\rm log}Q_{\rm int}=\sum\limits_{i=0}^{6} a_i({\rm log}T)^i.
\label{Qint}
\end{eqnarray}

In order to get the best reproduction of the directly computed values,
the fit had to be performed in two separate temperature ranges.
The first range is $0-200$ K, the other is $201-4,000$ K.
These fits can reproduce the values of log$Q$  reasonably accurately,
within 0.1\% in either region.
Nevertheless, to take full advantage of the high accuracy of the present
thermochemical reults for C$_2$, the numerical results of the
Supplementary Information should be used. This supplementary information
also contains the transitions file which forms the input for MARVEL,
and which can be augmented with any future spectroscopic data on C$_2$ and
rerun, and the associated energies file which is the output from the
MARVEL run.

\renewcommand{\baselinestretch}{1.0}
\begin{deluxetable}{crr}
\tablecaption{Coefficients of the fit, see Eq. (3), to the internal partition function of $^{12}$C$_2$.}\label{tab:Qcoeff}
\tablewidth{0pt}
\tablehead{
\colhead{Coefficient }    & \colhead{0 -- 200 K}  & \colhead{201 -- 4000 K} }
\startdata
$a_0$  &  3.6577362306 &  186.8558092069   \\
$a_1$  & $-$7.0625294443 & $-$139.9034556834   \\
$a_2$  &  5.4057108902 &   42.4445304374   \\
$a_3$  & $-$2.0744319852 &   $-$6.6283715970   \\
$a_4$  &  0.4447612738 &    0.5612382751   \\
$a_5$  & $-$0.0504276952 &   $-$0.0241572167   \\
$a_6$  &  0.0023610646 &   0.00040521942   \\
\enddata
\end{deluxetable}

\begin{deluxetable}{rrrrlcl}%[!htbp]
\tablecaption{Temperature-dependent thermochemical data for $^{12}$C$_2$.
The uncertainties associated with the data are given in parentheses.}\label{tab:Q}
\tablewidth{0pt}
\tablehead{
\colhead{$T$/K}&\colhead{$Q_{\rm int}(T)$}&\colhead{$Q'_{\rm int}(T)$}&\colhead{$Q''_{\rm int}(T)$}&\colhead{$C_{\rm p}(T)$ / J~K$^{-1}$~mol$^{-1}$}}
\startdata
100.0    &	19.3786(1)    &  19.3839(8)     &  40.242(8)    &  29.7333(2) \\
200.0    &	41.6629(1)    &  55.1971(7)     &  170.867(15)   &  40.2915(2) \\
300.0    &	78.0529(1)    &  137.7708(7)    &  453.195(28)   &  43.1578(2) \\
400.0    &	133.6964(1)   &  260.3942(7)    &  812.605(39)   &  39.7816(2) \\
500.0    &	207.5463(1)   &  411.2436(7)    &  1225.949(48)  &  37.2546(2) \\
600.0    &	297.5930(1)   &  585.6218(7)    &  1700.047(59)  &  36.0862(2) \\
700.0    &	402.3796(1)   &  782.8055(7)    &  2244.006(70)  &  35.6865(2) \\
800.0    &	521.0433(1)   &  1003.3874(7)   &  2863.969(87)  &  35.6538(2) \\
900.0    &	653.1204(1)   &  1248.2596(7)   &  3564.19(15)  &  35.7888(2) \\
1000.0   &	798.3910(1)   &  1518.3519(7)   &  4348.50(38)  &  36.0006(2) \\
1100.0   &	956.7878(1)   &  1814.6138(7)   &  5221.04(92)  &  36.2499(2) \\
1200.0   &	1128.3464(1)  &  2138.0495(7)   &  6186.5(19)  &  36.5204(3) \\
1300.0   &	1313.1770(1)  &  2489.7472(7)   &  7250.5(37)  &  36.8053(5) \\
1400.0   &	1511.4500(2)  &  2870.8933(7)   &  8418.9(65)  &  37.1015(7) \\
1500.0   &	1723.3857(2)  &  3282.7725(13)   &  9698(10)  &  37.4068(10) \\
1600.0   &	1949.2475(3)  &  3726.7596(19)   &  11095(15) &  37.7195(14) \\
1700.0   &	2189.3369(4)  &  4204.3059(28)   &  12616(22) &  38.0375(17) \\
1800.0   &	2443.9883(6)  &  4716.9251(38)   &  14269(30) &  38.3585(21) \\
1900.0   &	2713.5652(8)  &  5266.1783(51)   &  16060(41) &  38.6803(25) \\
2000.0   &	2998.4562(11)  &  5853.6614(65)   &  17996(53) &  39.0006(29) \\
2100.0   &	3299.0714(15)  &  6480.9933(83)   &  20084(66) &  39.3173(33) \\
2200.0   &	3615.8393(19)  &  7149.806(11)   &  22332(81) &  39.6285(36) \\
2300.0   &	3949.2042(24)  &  7861.740(17)   &  24744(98) &  39.9326(40) \\
2400.0   &	4299.6233(32)  &  8618.434(34)   &  27329(116) &  40.2281(43) \\
2500.0   &	4667.5647(49)  &  9421.521(74)   &  30092(135) &  40.5139(50) \\
2600.0   &	5053.5054(88)  &  10272.63(15)  &  33039(156) &  40.7893(63) \\
2700.0   &	5457.929(17)  &  11173.38(32)  &  36178(178) &  41.0535(92) \\
2800.0   &	5881.327(33)  &  12125.37(63)  &  39513(201) &  41.306(14) \\
2900.0   &	6324.193(64)  &  13130.2(12)  &  43051(225) &  41.546(23) \\
3000.0   &	6787.02(11)  &  14189.4(21)  &  46798(251) &  41.775(36) \\
3100.0   &	7270.32(21)  &  15304.7(36)  &  50760(280) &  41.991(56) \\
3200.0   &	7774.59(36)  &  16477.4(60)  &  54941(316) &  42.195(83) \\
3300.0   &	8300.33(61)  &  17709.2(98)  &  59348(363) &  42.38(12) \\
3400.0   &	8848.04(98)  &  19001(15)  &  63986(429) &  42.56(17) \\
3500.0   &	9418.2(15)  &  20355(23)  &  68859(528) &  42.73(23) \\
3600.0   &	10011.4(23) &  21773(35)  &  73971(673) &  42.89(32) \\
3700.0   &	10628.0(35) &  23256(52)  &  79328(882) &  43.03(42) \\
3800.0   &	11268.6(52) &  24805(74)  &  84933(1172) &  43.16(56) \\
3900.0   &	11933.7(75) &  26421(105)  &  90789(1562) &  43.28(71) \\
4000.0   &	12624(11) &  28106(146)  &  96899(2076) &  43.39(91) \\
\enddata

%\tablecomments{}
\end{deluxetable}

%\newpage
\renewcommand{\baselinestretch}{1.25}
\section{Summary}

This study utilizes the MARVEL technique to accurately determine
close to 6,000 experimental rovibronic energies of $^{12}$C$_2$
for six singlet, six triplet, and two quintet electronic states,
including the eight lowest valence states which gives coverage up to 35~000 \cm.
We survey all available laboratory high-resolution spectroscopic studies
to provide input data for this process, resulting in 23,343
transitions connecting the 14 electronic states.
While there are many spectroscopic studies available,
in fact 42 were analyzed to yield the transitions analyzed, and there has been
significant recent activity including the identification of several
new band systems, there are also surprising gaps.
For example, there is a detailed, fully rovibronically-resolved study of the
Deslandres--d'Azambuja (\Cstate $-$ \Astate) band  system
for $^{13}$C$_2$ \citep{85AnBoPe.C2}, but
even a century after the original observation of this band  \citep{1905DeDa.C2}
there is no available high-resolution study for $^{12}$C$_2$.

The recent observation of singlet--triplet intercombination bands by \citet{15ChKaBeTa.C2}
helped us to achieve to link all rovibronic levels of $^{12}$C$_2$ into a single
huge component within its experimental spectroscopic network;
thus, individual intercombination lines can now be predicted accurately
using the results of our study.
This is a significant step towrd the astronomical detection of these
transitions \citep{86LeRo}.
To further aid this work and other astronomical studies involving C$_2$,
a full rovibronic line list for $^{12}$C$_2$
is currently being constructed using the variational code {\sc Duo}
by \cite{jtexoC2}, as part of the ExoMol project \citep{jt528}.

As to now, the full set of MARVEL results comprising a file of validated
transition frequencies and a file containing the  resulting rovibronic
energy levels are given in the Supplementary Information to this paper.
The highly accurate but limited set of experimental (MARVEL) energy levels
augmented with the much less accurate but much more complete set of {\sc Duo}
energy levels has been used to compute ideal-gas thermochemical functions
for $^{12}$C$_2$ up to 4,000 K.
The accuracy of the partition function is better than 0.1\% even at the
highest temperatures, considerably exceeding the accuracy of all previous
studies.
This assures that the accuracy of the present isobaric heat capacity of
$^{12}$C$_2$ is significantly better than that of any previous study.

\section*{Acknowledgement}

We thank Robert Curl, John Hardwick, Michael Heaven, and Jian Tang for supplying
the (unpublished) data from their spectroscopic experiments, and
Tatiana Petrova and Peter Radi for comments on their data.
Peter Radi is also thanked for his comments on the manuscript.
This work has received support from
the European Research Council under Advanced Investigator Project 267219,
and the Scientific Research Fund of Hungary (grant OTKA NK83583).
Collaboration of the UCL and ELTE groups has greatly benefited from
the support of two COST actions, CoDECS (CM1002) and MOLIM (CM1405).
Some funding was provided by the NASA Laboratory Astrophysics Program.

%\newpage
\bibliographystyle{apj}
%\bibliography{journals_astro,C2,jtj,gen,partition}

\end{document}